\documentclass[aps,prb,%draft,
  twocolumn,showpacs,
  superscriptaddress,groupedaddress,floatfix]{revtex4-1}
\usepackage{amsmath,amssymb,graphicx,mathtools}
\usepackage[dvipsnames]{xcolor}

\begin{document}
\title{Dynamics of almost strong edge modes 
in spin chains away from integrability}
\author{Daniel J. Yates$^{1}$} \author{Alexander G. Abanov$^{2,3}$}
\author{Aditi Mitra$^{1}$} \affiliation{$^{1}$Center for Quantum
  Phenomena, Department of Physics, New York University, 726 Broadway,
  New York, NY, 10003, USA\\ $^{2}$Simons Center for Geometry and
  Physics, Stony Brook, NY 11794, USA\\ $^{3}$Department of Physics
  and Astronomy, Stony Brook University, Stony Brook, NY 11794, USA}

\date{\today}
\begin{abstract}
  Results are presented for the dynamics of an almost strong edge mode
  which is the quasi-stable Majorana edge mode occurring in
  non-integrable spin chains. The dynamics of the edge mode is studied
  using exact diagonalization, and compared with time-evolution with
  respect to an effective semi-infinite model in Krylov space obtained
  from the recursion method.  The effective Krylov Hamiltonian is
  found to resemble a spatially inhomogeneous SSH model where the
  hopping amplitude increases linearly with distance into the bulk,
  typical of thermalizing systems, but also has a staggered or
  dimerized structure superimposed on it. The non-perturbatively long
  lifetime of the edge mode is shown to be due to this staggered
  structure which diminishes the effectiveness of the linearly growing
  hopping amplitude. On taking the continuum limit of the Krylov
  Hamiltonian, the edge mode is found to be equivalent to the
  quasi-stable mode of a Dirac Hamiltonian on a half line, with a mass
  which is non-zero over a finite distance, before terminating into a
  gapless metallic bulk. The analytic estimates are found to be in
  good agreement with the numerically obtained lifetimes of the edge
  mode.
\end{abstract}
\maketitle

%%%%%%%%%%%%%%%%%%%%%%%%%%%%%%%%%%%%%%%%%%%%%%%%%%%%%%%%%%%%%%%%%%%%%%%%%%%%%%%%%%%%%%%%%%%%%
%%%%%%%%%%%%%%%%%%%%%%%%%%%%%%%%%%%%%%%%%%%%%%%%%%%%%%%%%%%%%%%%%%%%%%%%%%%%%%%%%%%%%%%%%%%%%
%%%%%%%%%%%%%%%%%%%%%%%%%%%%%%%%%%%%%%%%%%%%%%%%%%%%%%%%%%%%%%%%%%%%%%%%%%%%%%%%%%%%%%%%%%%%%
\section{Introduction}

Topological systems hosting Majorana zero modes have seen extensive
research efforts over the past two decades, both in theory and in
experiment, due to their potential to provide nearly error-free
quantum computation through the braiding manipulations of non-local
fermions \cite{NayakRMP08,Alicea12,Beenaker13,Freedman15}. In
particular, spinless superconductors in one dimension (1D) are a
potential host for these operators \cite{Kitaev01}, and there are a
growing number of experimental platforms to realize them
\cite{Leo12,Marcus16,Yazdani14,Yazdani19,Kraus12,Loss13,Liu18}.

However, systems that host topological degrees of freedom at the
boundary are often only well understood in the free limit
\cite{Bernevigbook,Hasan10,Ryu10,QiRev11}, or in some cases the
interacting but zero temperature limit \cite{Wen17, Hosho15}.  At
finite temperatures, as one turns on interactions, the edge modes will
have finite lifetimes for generic systems. One would expect that
temperatures that are small as compared to the single particle
topological gap to be the most experimentally relevant. Yet, many
intriguing examples are beginning to emerge, where edge modes are
stable for long times even at temperatures higher than this bulk
single-particle gap \cite{Nayak17,
  Fendley17,Parker19,Yao19,Pollmann20,Yates20}.

Floquet systems provide new avenues to realize topological edge modes
and new topological phases with no static analogs
\cite{Kitagawa10,Rudner13,Zoller11,Yao17,
  Yates17,Yates18,Liu19,Sondhi20}. As energy is no longer conserved,
and is in-fact being pumped into the system leading to heating
\cite{Lazarides15,Hyungwon14,DAlessio14,Ponte15,Haldar18}, infinite
temperature studies of the lifetime of edge modes in the context of
Floquet systems, are also necessary.

In the face of non-integrable, non-equilibrium dynamics, there are
largely two active fields of research that study ways to (somewhat)
rein-in ergodicity: many-body localization \cite{mblrev} and
prethermalization \cite{Abanin15, Mori16, Abanin17,Abanin17b}.  This
work will focus on clean systems and is therefore unrelated to
many-body localization. Additionally, the robustness of the edge modes
that will be discussed is not related to
prethermalization in the usual sense as the dynamics associated with
bulk operators is fully thermalizing. Yet, even with infinite
temperatures, moderate interaction strengths, and no disorder, we
present situations where edge operators that have an overlap with the
topological edge mode in the free limit, can survive for times much
longer than bulk thermalization times.

A central object in the study of long lived edge modes has been the
concept of the strong zero mode (SM)
\cite{Kitaev01,Fendley12,Fendley14,Fendley16} and the almost strong
zero mode (ASM)\cite{Fendley17,Nayak17}. The SM usually can be thought
of as many-body generalizations of the Majorana zero modes, however
there has been an example of a SM that exists for an
integrable-interacting system \cite{Fendley16}.  The ASM is
essentially what becomes of the SM when integrability is broken and it
is this quantity whose lifetime is studied
\cite{Fendley17,Parker19,Yao19,Yates20}.

Let us note that the existence of a SM is a much stronger
statement than the existence of an edge zero mode as the former is a
statement for the entire spectrum, not just the ground state. The
existence of a SM implies the existence of an edge zero mode, but
not vice versa. However whether the edge zero mode is a topological
edge mode is a more subtle question.  To the best of our knowledge
there is no general statement relating the two. However, in some
known examples with a SM, the edge zero mode can be interpreted as a
topological edge mode after possibly an additional Jordan-Wigner
like non-local transformation \cite{Kitaev01}.

Solving the dynamics of ASMs has only been possible through exact
diagonalization (ED) calculations, with the lifetime of the edge
operator found to be non-perturbative in the integrability breaking
parameter\cite{Nayak17}.  However, lifetimes extracted numerically are
plagued with system size dependencies, and an approach valid in the
thermodynamic limit is needed.

We recently showed a route to estimating the lifetime of
ASMs\cite{Yates20}, utilizing the recursion method \cite{Recbook,
  Altman19}. This method maps the non-integrable dynamics of the
Heisenberg equations of motion for the operator of interest, onto that
of a free particle whose dynamics is governed by a 1D, semi-infinite,
nearest-neighbor, tight-binding model, also known as the Krylov
Hamiltonian.  Our previous study \cite{Yates20} suggested a new
interpretation of the slow dynamics of the ASM as the result of the
presence of an approximate topological edge state in the fictitious
lattice of the recursion method.  This approximate edge state is
similar to that of a Su-Schrieffer-Heeger (SSH) \cite{SSH79,SSH80}
edge mode, but is only quasi-stationary as the state eventually
becomes non-normalizable in the infinite bulk, implying an overlap
with bulk states that causes the mode to eventually decay.

This work further builds on our previous work \cite{Yates20} in the
following ways.  A toy model is constructed from the numerically
obtained parameters of the fictitious lattice, and the edge mode
operators of this toy model are discussed. Moreover analytic
expressions for the lifetime of the edge mode are derived from the toy
model, and compared with the numerically obtained lifetimes.

The paper is organized as follows. In Section \ref{model} the model is
introduced, the SM and ASM are defined, and the recursion method
outlined. In Section \ref{toy}, the toy model and some simple
variations of it are introduced, and their edge modes discussed.  In
addition, the parameters of the toy model are explicitly extracted
from the numerical data.  In Section \ref{Lanc}, the continuum limit
of the toy model is derived, and an analytic estimate for the lifetime
of the ASM is obtained. Following this, in Section
\ref{sec_green_setup}, the toy model is solved, without making the
continuum approximation, and an analytic estimate for the lifetime of
the ASM is obtained.  Comparison between the dynamics from the
discrete toy model and the ED dynamics are presented in Section
\ref{comp}, and we present our conclusions in Section
\ref{conclu}. Some details are relegated to the appendices.

%%%%%%%%%%%%%%%%%%%%%%%%%%%%%%%%%%%%%%%%%%%%%%%%%%%%%%%%%%%%%%%%%%%%%%%%%%%%%%%%%%%%%%%%%%%%%
%%%%%%%%%%%%%%%%%%%%%%%%%%%%%%%%%%%%%%%%%%%%%%%%%%%%%%%%%%%%%%%%%%%%%%%%%%%%%%%%%%%%%%%%%%%%%
%%%%%%%%%%%%%%%%%%%%%%%%%%%%%%%%%%%%%%%%%%%%%%%%%%%%%%%%%%%%%%%%%%%%%%%%%%%%%%%%%%%%%%%%%%%%%
%%%%%%%%%%%%%%%%%%%%%%%%%%%%
\section{Hamiltonian, Strong zero mode, Almost strong zero mode, and Recursion Method} \label{model}
%%%%%%%%%%%%%%%%%%%%%%%%%%%%

We study the \(XYZ\) spin 1/2 chain Hamiltonian with a transverse
magnetic field $g$,
\begin{align}
  H &= \sum_i \biggl[J\left( \frac{1 + \gamma}{2}\right)
    \sigma_i^x \sigma_{i+1}^x
    \left.+ J \left(\frac{1-\gamma}{2} \right)
    \sigma_i^y \sigma_{i+1}^y \right. \nonumber  \\
   &  \qquad + J_z \sigma_i^z \sigma_{i+1}^z + g \sigma_i^z\biggr]\,.
 \label{H}
\end{align}
We briefly discuss some limiting forms of the above model. For $J_z =0
$, and after a Jordan-Wigner transformation \cite{Lieb61,Mattis64},
the model maps to free fermions. For this free case, $\gamma = 1$
corresponds to the transverse field Ising model, \cite{Sachdevbook}
and is also equivalent to the Kitaev chain when written in the
Majorana representation \cite{Kitaev01}.

For \(J_z\neq 0\), and in the Majorana representation, the model
corresponds to a chain with nearest-neighbor interactions of strength
\(J_z\), and a superconducting gap of strength \(\gamma\).  When $J_z
\ne 0,g = 0$ the model is interacting but integrable. In contrast, for
$J_z \ne 0, g \ne 0$, the system is nonintegrable.  We set $J= 1$
throughout this paper, and denote the length of the chain by \(L\).
We will be interested in the nonintegrable case of \(J_z \neq 0, g\neq
0\).

A convenient starting point is to consider $J_z = 0, \gamma \ne 0 $
when \(H\) is similar to the Kitaev chain with a general
superconducting gap \(\gamma\), and the model hosts topological edge
states at the boundary. For these parameters, the system falls under
class D of the Altland-Zirnbauer classification scheme
\cite{AZ97,Ryu10,Fidk11} that is characterized by a discrete
\(\mathbf{Z}_2\) symmetry corresponding to fermion parity. The
symmetry is manifest even with interactions. This is evident through
the operator
\begin{align}\label{Dded}
    D = \sigma_1^z \dots \sigma_L^z,
\end{align}
which commutes with the Hamiltonian \eqref{H} for all values of
coupling constants.

%%%%%%%%%%%%%%%%%%%%%%%%%%%%%%%%%%%%%%%%%%%%%%%%%%%%%%%%%%%%%%%%%%%%%%%%%%%%%%%%%%%%%%%%%%%%%
%%%%%%%%%%%%%%%%%%%%%%%%%%%%%%%%%%%%%%%%%%%%%%%%%%%%%%%%%%%%%%%%%%%%%%%%%%%%%%%%%%%%%%%%%%%%%
%%%%%%%%%%%%%%%%%%%%%%%%%%%%%%%%%%%%%%%%%%%%%%%%%%%%%%%%%%%%%%%%%%%%%%%%%%%%%%%%%%%%%%%%%%%%%
%%%%%%%%%%%%%%%%%%%%%%%%%%%%
\subsection{Strong zero mode (SM)} 

A key concept that readily generalizes free topological edge states to
the interacting case, is that of Strong zero modes (SM)
\cite{Fendley16,Nayak17,Fendley17}. A SM is defined as an operator
$\Psi$, which commutes with $H$ in the thermodynamic limit, $[H,\Psi]
\rightarrow 0, L \rightarrow \infty$, anti-commutes with the global
symmetry $\{D, \Psi\} = 0$, and is normalizable i.e., \(\Psi^2 =
O(1)\).  The SM are a statement about the full spectrum of $H$ rather
than particularities of the ground state.  In particular, the
existence of a SM implies that the full spectrum of $H$ is at least
doubly degenerate, corresponding to the two different symmetry
sectors.

In the free limit, the single-particle topological edge operators are
precisely SMs, and in this limit when \(g=0\), the SM is trivially
\(\Psi = \sigma^x_1\) or, via a Jordan-Wigner transformation, \(\Psi=
a_1\), where \(a_1\) is the Majorana mode on the first site. In the
Ising limit of \(g \neq 0, \gamma =1\) the SM has been discussed in
Ref.~\onlinecite{Kitaev01,Fendley16}, while for \(g\neq 0, \gamma\neq
1\), the SM was constructed in Ref.~\onlinecite{Yates20}.  The
interacting-integrable XYZ model (\(g=0\)) was also shown to host a SM
localized at the edge of a semi-infinite system \cite{Fendley16}. As
these operators commute with $H$ in the large system-size limit, their
dynamics is trivial. For small system sizes on the other hand, the SM
can decay by tunneling across the wire, acquiring a lifetime that is
exponential in system size\cite{Fendley16,Fidk11a}. This lifetime can
be estimated using perturbative arguments~\cite{Yates20}.

%%%%%%%%%%%%%%%%%%%%%%%%%%%%%%%%%%%%%%%%%%%%%%%%%%%%%%%%%%%%%%%%%%%%%%%%%%%%%%%%%%%%%%%%%%%%%
%%%%%%%%%%%%%%%%%%%%%%%%%%%%%%%%%%%%%%%%%%%%%%%%%%%%%%%%%%%%%%%%%%%%%%%%%%%%%%%%%%%%%%%%%%%%%
%%%%%%%%%%%%%%%%%%%%%%%%%%%%%%%%%%%%%%%%%%%%%%%%%%%%%%%%%%%%%%%%%%%%%%%%%%%%%%%%%%%%%%%%%%%%%
%%%%%%%%%%%%%%%%%%%%%%%%%%%%
\subsection{Almost strong zero mode (ASM)}

As the parameters are tuned away from the integrable point, the
operator $\Psi$ will no longer be a SM, with the commutation of $H$
failing to drop off as the system-size is increased en-route to the
thermodynamic limit. In this case, where the commutator is very small
albeit non-zero in the thermodynamic limit, the operators are called
ASMs, and their dynamics is in between that of the trivial dynamics of
the exact SM and the featureless dynamics of an infinite temperature
non-integrable system
\cite{Altman19,Gorsky19,Barbon19,Avdoshkin19}. In particular, the ASM
is typically characterized by long lifetimes, despite strong
interaction strengths and infinite temperatures
\cite{Fendley17,Yates20}.

One way to study the dynamics of a (A)SM, which is also particularly
well suited to cases where analytic expressions for the SM are
unavailable, is to consider the edge auto-correlation function
$A_\infty$, defined as,
\begin{equation}\label{Adef}
  A_\infty(t) =
  \frac{1}{2^L} \text{Tr} \left[ \sigma_1^x(t) \sigma_1^x(0)\right].
\end{equation}
This quantity will act as a good measure of the lifetime of the ASM as
long as $\text{Tr}[\Psi \sigma_1^x ] \sim O(1)$. $A_\infty$ measures
the lifetime of a Majorana mode to remain on the first site, or
equivalently it measures the edge spin coherence as one time-evolves a
generic state. The latter is a state that is far from being any exact
eigenstate of \(H\) and therefore has overlap over a large portion of
the spectrum of \(H\).

\begin{figure}
  \includegraphics[width = .49\textwidth]{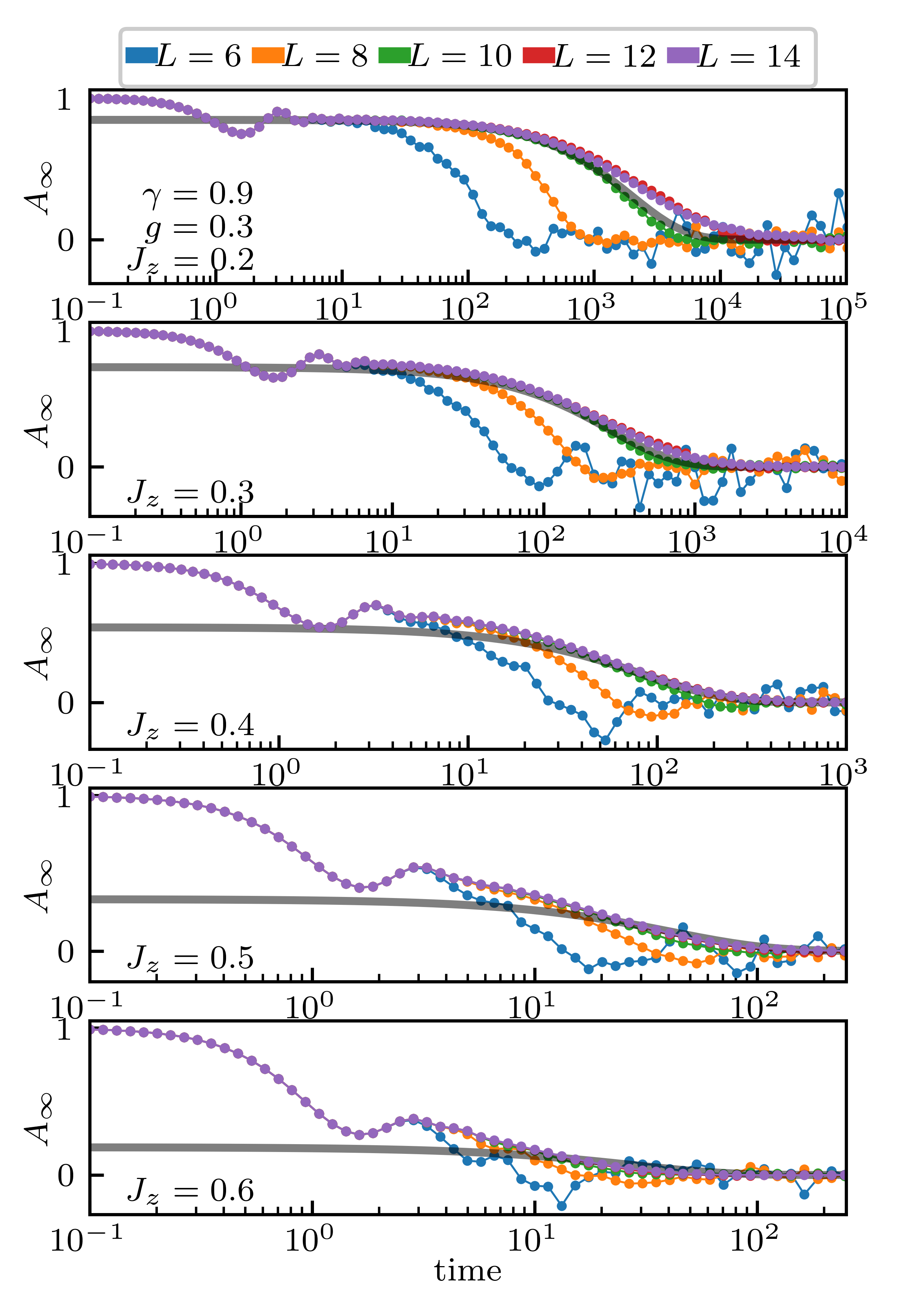}
  \caption{
  $A_\infty$ for system lengths $L = 6,8,10,12,14$, $\gamma =
    .9$, $g = .3$. From top to the bottom panel $J_z =
    .2,.3,.4,.5,.6$.  The dotted data sets are the result of ED
    calculations.  The solid black curve is the result of
    approximating $A_\infty$ 
    by Eq.~\eqref{eq_ainf_approx1},
    with $N = 400$, where $b_n$ used in this computation,
    are generated by the Lanczos algorithm 
    Eq.~\eqref{eq_lanczos}, for $L = 12$, and are shown in 
    Fig.~\ref{fig_bs}. Note the different ranges on the \(x\)-axes.}
  \label{fig_ainfs}
\end{figure}

Examples of $A_\infty(t)$ for a range of $J_z$ are shown in
Fig.~\ref{fig_ainfs}.  The lifetime of the ASM is sensitive to small
system sizes. In Fig.~\ref{fig_ainfs}, the lifetime increases as the
system size increases. At some point, a large enough system size is
reached where the lifetime becomes system size independent, or
saturates, allowing ED to be representative of the thermodynamic limit
for this quantity. In Fig.~\ref{fig_ainfs}, this limit is reached for
all parameter values.

There is a growing series of studies that predict a lifetime
dependence near the Ising limit \(\gamma \sim 1\), in terms of the
integrability breaking term $J_z$, of the form $\sim e^{1/J_z}$ up to
logarithmic corrections \cite{Nayak17,Parker19,Yao19, Yates20}.  This long
lifetime can be qualitatively predicted by using prethermal type
arguments.  In particular, the \(O(1)\) energy scale to flip the spin
at the boundary is highly off-resonant with the microscopic scale
\(J_z\) that breaks integrability since \(J_z\ll 1\). Thus, the long
lifetime arises from counting the number of microscopic processes
\(1/J_z\) needed to flip the boundary spin. However as mentioned
earlier, the dynamics of bulk operators are fully thermalizing, and
this prethermal argument is only valid for the edge operator.

Long lived ASMs have also been demonstrated when the prethermal
concept is clearly not applicable, namely when a large energy scale
separation is absent in the free case because an external drive in a
Floquet setting is not highly off-resonant with the
system\cite{Yates19}.  Further, long-lived ASM analogs of Floquet
\(\pi\)-modes, which require drives resonant with the single-particle
band have been demonstrated even for the clean system with
interactions\cite{Yates19}.

A more general perspective to the non-perturbative origin of the
lifetime of the ASM, not dependent on prethermal physics, was
presented in Ref.~\onlinecite{Yates20}. Here, the dynamics of the edge
operator was mapped to that of a single-particle semi-infinite chain,
by means of the recursion method \cite{Recbook, Altman19}. The
semi-infinite chain obtained this way was found to have topological
features similar to that of the SSH model, and therefore this mapping
allowed one to show that the slow dynamics of the ASM was a result of
a long lived edge mode of a generalized or perturbed SSH model.

The goal of the current paper is to further build on the results of
Ref.~\onlinecite{Yates20}. A more detailed numerical study of the
dynamics of the ASM will be supplemented by analytic estimates for the
lifetime. The latter will be obtained by exploiting the mapping to a
free system to construct suitable toy models. The quasi-stable edge
modes of the toy model will be solved for.

\subsection{Lanczos algorithm and the Krylov Hamiltonian} %\label{Lanc}
The time evolution of an operator in a generic integrable or
non-integrable system can be mapped to single particle dynamics on a
semi-infinite chain \cite{Recbook}. In this section we outline this
method.  The exponential complexity of solving the dynamics enters
into the calculation of the hopping parameters on this chain which we
denote by $b_n$.

The Heisenberg time-evolution operator can be written as
\begin{equation}
  e^{i H t} O e^{-i H t} = \sum_{n = 0}^\infty \frac{(i t)^n}{n!} \mathcal{L}^n O,
\end{equation}
where we define
\begin{equation}
  \mathcal{L}O = [H,O].
\end{equation}
To employ the Lanczos algorithm, we will recast operator dynamics into
vector dynamics by defining $|O) = O$.  Since we are concerned with
infinite temperature quantities, we have an unambiguous choice for an
inner product on the level of the operators,
\begin{equation}
  (A|B) = \frac{1}{2^L} \text{Tr} \left[ A^\dagger B \right].
\end{equation}

The Lanczos algorithm iteratively finds the operator basis that
tri-diagonalizes $\mathcal{L}$. We begin with the seed ``state'',
$|O_1)$, and let $\mathcal{L}|O_1) = b_1 |O_2)$, where $b_1 =
\sqrt{|\mathcal{L}|O_1)|^2}$.  The recursive definition for the basis
operators $|O_{n\ge 2})$ is,
\begin{equation}
  \mathcal{L} |O_n) = b_{n} |O_{n+1}) + b_{n-1} |O_{n-1}),
  \label{eq_lanczos}
\end{equation}
where we define $b_n = \sqrt{|\mathcal{L}|O_{n})|^2}$.  It is
straightforward algebra to check that the above procedure will
iteratively find basis operators that yield a $\mathcal{L}$ which is
tri-diagonal, and of the following form,
\begin{equation}\label{Lmat}
  \mathcal{L} =
  \begin{pmatrix}
    & b_1 &     &     & \\
    b_1 &     & b_2 &     & \\
    & b_2 &     & \ddots & \\
    &     & \ddots&   &
  \end{pmatrix}.
\end{equation} 
This basis spanned by $|O_n)$ lies within the Krylov sub-space of
$\mathcal{L}$ and $|O_1)$. We refer to this tri-diagonal matrix as the
Krylov Hamiltonian.

An important aspect of this technique, often overlooked when
discussing chaos, is that the values of $b_n$ are highly dependent on
the choice of seed operator. Further, outside of special cases, namely
a Hamiltonian that is free, the exact solution to the operation
$\mathcal{L}|O_n)$ will require ED, or similar methods with equivalent
costs. This method does not escape the rapidly growing exponential
wall of complexity.  In cases where the calculation of all $b_n$ are
possible, the above algorithm will return a value of $b_{\text{end}} =
0$. Knowledge of the full set of $b_n$ results in full knowledge of
the dynamics of the seed operator.
  
For free systems, the operation $\mathcal{L}|O_n)$ can be efficiently
solved when in the Majorana basis. If the starting operator is a
single Majorana then the dimension of the Krylov-subspace of that
operator will scale as $2L$, as free system dynamics can only mix the
individual Majoranas among themselves.
  
Outside of free problems, the size of the full set of $|O_n)$ will be
large. For example, a system size of \(L\) will have $\sim 2^{2L}$
possible basis operators. For all intents and purposes we treat
$\mathcal{L}$ as a semi-infinite chain.  If the number of solved $b_n$
is insufficient for the quantity of interest, the typical approach is
to supplement the known set with approximate $b_{n}$ that are
calculated based off of trends established among the known hoppings.

Starting with the seed state $|O_1) = |\sigma_1^x)$, we can recast
$A_{\infty}$ into an equivalent form,
\begin{equation}
  A_{\infty}(t) = (e^{i \mathcal{L} t})_{1,1}.
\end{equation}
Now, following the above discussion, the dynamics of $A_{\infty}$ has
been transformed into that of a semi-infinite single-particle problem.
The details of the semi-infinite chain will be discussed in subsequent
sections. Ref.~\onlinecite{Yates20} showed that the slow dynamics of
$A_\infty$ is a result of topological modes residing at the left
boundary (origin) of the Krylov wire.  Ref.~\onlinecite{Yates20} also
showed how the parameters of the Krylov wire change for different
choices of edge operators such as $|O_1) = |\sigma_1^{y,z})$.

\section{General structure of \(b_n\) and the toy model.} \label{toy}
In this section we will discuss the general structure of the Krylov
wire in Eq.~\eqref{Lmat} parameterized by the hoppings \(b_n\), and
use our observations to motivate some toy models.

\subsection{Discussion of $b_n$} \label{sec_3A}

\begin{figure*}
  \includegraphics[width = .99\textwidth]{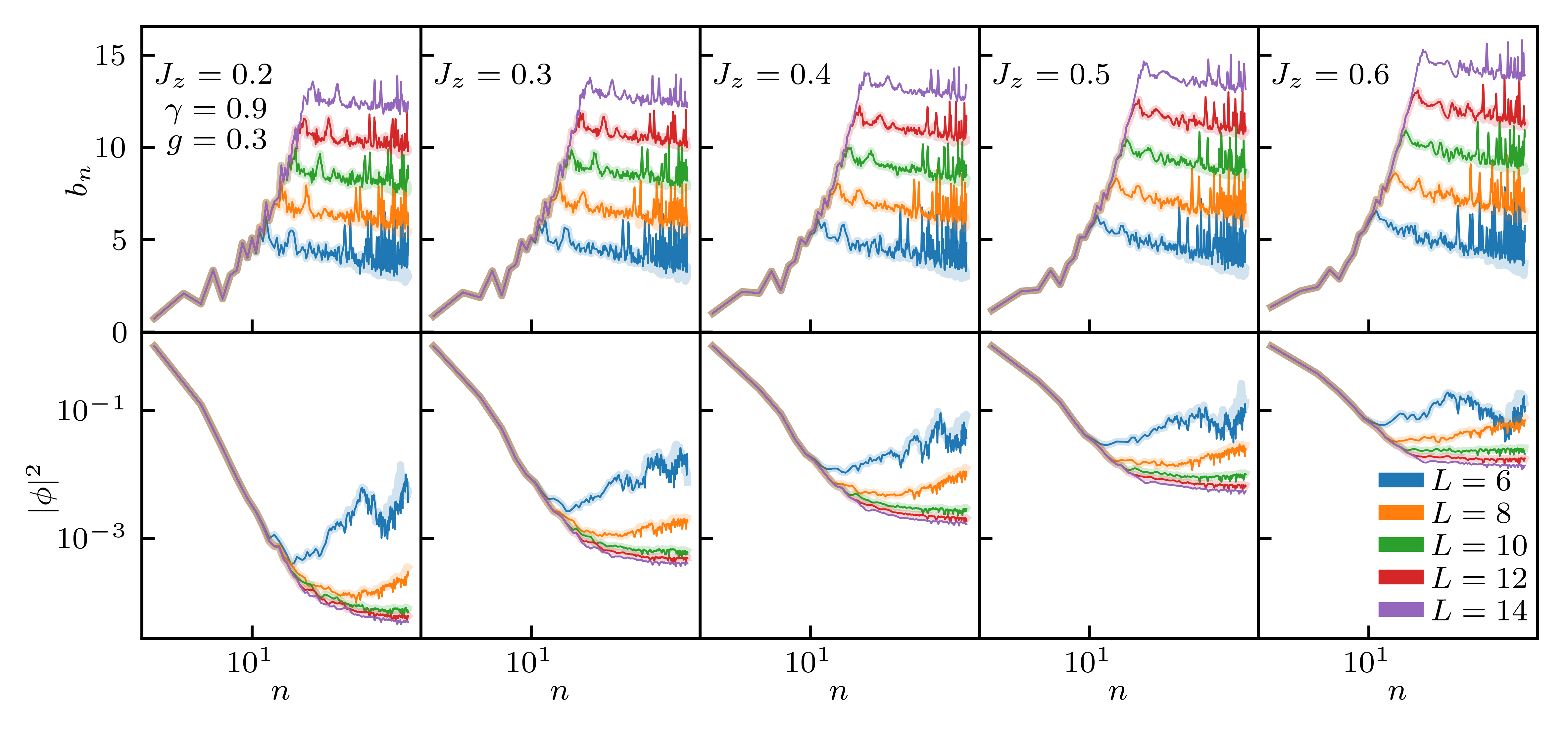}
  \caption{Top panel: $b_n$ plotted for system lengths $L =
    6,8,10,12,14$, $\gamma = .9$, $g = .3$. Left to right, $J_z =
    .2,.3,.4,.5,.6$. The sharp, thin data-sets correspond to the
    Lanczos algorithm, Eq.~\eqref{eq_lanczos}.  The translucent, thick
    data-sets for $L = 6$-$12$, correspond to the Gram-Schmidt
    orthogonalized Lanczos algorithm discussed in Appendix
    \ref{bnerror}.  Bottom: The mod-square of the wave-function
    $|\phi|^2$ of the edge state of the Krylov lattice plotted for the
    same parameters. The thin, sharp data set corresponds to
    calculating $\phi$ for the thin sharp $b_n$ of the top row, with
    analogous statements for the translucent, thick data set.  The
    $\phi$ are not normalized, with $\phi_1 = 1$ for all data sets.  }
  \label{fig_bs}
\end{figure*}

Fig.~\ref{fig_bs} shows the first 400 $b_n$ calculated for the same
parameters as those in Fig.~\ref{fig_ainfs}.  There are three main
aspects to the $b_n$, the roughly linear ramp for small $n$, the
staggering or dimerization of the $b_n$ superimposed on the ramp, and
the system size dependent ``plateau'' or flat region after the linear
ramp ends. Note that we will use the words staggering and dimerization
interchangeably to denote the even-odd pattern of the \(b_n\) typical
of an SSH model~\cite{SSH79,SSH80}.

The linear ramp is expected for non-integrable systems
\cite{Altman19}, with the slope increasing with \(J_z\)
\cite{Yates20}. In contrast, the staggering becomes more pronounced as
$J_z$ is decreased, and coincides with the longer-lived ASM
\cite{Yates20}. The $b_n$ resemble an SSH model with a sign of the
dimerization which is topological, but with a linear slope
superimposed on the hoppings. We expand on these points further when
we develop the toy model.

Fig.~\ref{fig_bs} also shows that the height of the plateau does not
experience a saturation in system size, $L$, even though $A_\infty$ in
Fig.~\ref{fig_ainfs} does saturate with system size. This implies that
the lifetime of $A_\infty$ must be independent of plateau heights, and
rather, must be primarily dependent on the nature of staggering and
the linear ramp.

As emphasized earlier, many features depend on the initial seed
operator. Thus, while an overall slope is expected for all operators
in a non-integrable model, the nature of the dimerization, such as
its sign, and how long it persists into the bulk, will depend on
whether the operator has strong overlap with the ASM or whether an
operator has little overlap. For the latter, the autocorrelation
function will have a short lifetime~\cite{Yates20}.

One can supplement the naive Lanczos algorithm in
Eq.~\eqref{eq_lanczos} with an additional step of Gram-Schmidt
orthogonalization of the newly produced $|O_n)$, against $|O_{n'})$
for all $n'<n$. This ensures orthogonality and is necessary for
discussing the details of $b_{n \gtrapprox 100}$.  A comparison
between the Lanczos algorithm and the Gram-Schmidt orthogonalization
is presented in Appendix \ref{bnerror}.  However it is helpful to note
here that since the error between the two numerical schemes is
apparent only at large \(n\), for the times shown in
Fig.~\ref{fig_ainfs}, the approximate $A_{\infty}$ (discussed later,
c.f. Eq.~\eqref{eq_ainf_approx1}) sees minor changes when the
Gram-Schmidt is implemented.  Moreover, the Gram-Schmidt
orthogonalization requires too much computer resources to be performed
for $L=14$, and hence the corresponding data is absent in this paper.

The two different data sets, one from the Lanczos algorithm, and the
other from the Gram-Schmidt orthogonalization, for accessible $L$ and
different $J_z$, are shown in the top row of Fig.~\ref{fig_bs}.  As
mentioned above, the two data sets have perfect agreement for $n<100$
and qualitative agreement for the entire domain.  Outside of
Fig.~\ref{fig_ainfs} and Fig.~\ref{fig_bs}, we will concentrate on the
Gram-Schmidt orthogonalized data. Although this data set is available
only up to $L=12$, Fig.~\ref{fig_ainfs} shows that $L = 12$ is
sufficient to attain saturated i.e, system size independent lifetimes
for all shown $J_z$.

%%%%%%%%%%%%%%%%%%%%%%%%%%%%%%%%%%%%%%%%%%%%%%%%%%%%%%%%%%%%%%%%%%%%%%%%%%%%%%%%%%%%%%%%%%%%%
%%%%%%%%%%%%%%%%%%%%%%%%%%%%%%%%%%%%%%%%%%%%%%%%%%%%%%%%%%%%%%%%%%%%%%%%%%%%%%%%%%%%%%%%%%%%%
%%%%%%%%%%%%%%%%%%%%%%%%%%%%%%%%%%%%%%%%%%%%%%%%%%%%%%%%%%%%%%%%%%%%%%%%%%%%%%%%%%%%%%%%%%%%%
%%%%%%%%%%%%%%%%%%%%%%%%%%%%%%%%%%%%%%%
\subsection{Calculation of edge states of the Krylov lattice}

The \(b_n\)s in Fig.~\ref{fig_bs} suggest a topological SSH model with
a linear slope. Therefore the topological edge states of this
generalized SSH model become a likely candidate for the origin of the
slow dynamics of the ASM.  Motivated by this, we search for the form
of the edge states for generalized SSH models.  The calculation is the
same as in a regular SSH model with no slope.

We solve the eigenvalue equation $\mathcal{L}|\psi\rangle =
E|\psi\rangle$, which due to the tri-diagonal form of \(\mathcal{L}\)
in Eq.~\eqref{Lmat} can be rewritten into an iterative transfer matrix
form,
\begin{equation}
  \begin{pmatrix}
    \psi_{l+1} \\ \psi_{l+2}
  \end{pmatrix}
  =
  \begin{pmatrix}
    0 & 1\\
    -\frac{b_l}{b_{l+1}} & \frac{E}{b_{l+1}}
  \end{pmatrix}
  \begin{pmatrix}
    \psi_l \\ \psi_{l+1}
  \end{pmatrix}.
  \label{eq_edge_modes}
\end{equation}
Defining, 
$\phi_{2l-1} = \psi_{2l-1}$ and $\eta_{2l} = \psi_{2l}$, 
at $E= 0$ we have,
\begin{subequations}\label{eq_zero_modes}
\begin{align}
    \phi_{2l+1} &= -\frac{b_{2l-1}}{b_{2l}} \phi_{2l-1},\\
    \eta_{2l+2} &= -\frac{b_{2l}}{b_{2l+1}} \eta_{2l}.
\end{align}
\end{subequations}
Thus the even sites and odd sites decouple for \(E=0\).  The modes
with support on the odd (even) sites can be calculated by substituting
$\psi_{1(2)} = 1$ into the above equation.

As a point of orientation, note that for an ideal semi-infinite SSH
model extending from the origin to the right, $\phi$ is normalizable
and corresponds to the left boundary mode when in the topological
phase.  For this case, $b_1/b_2 = b_{2n-1}/b_{2n}$, and the normalized
$\phi$ is found to be ,
\begin{align}\label{pie}
\phi_{2l-1} = \sqrt{1-r^2} \,r^{l-1}; \;\; r=\frac{b_1}{b_2}.
\end{align}

Eq.~\eqref{eq_edge_modes} is completely general and $\phi,\eta$ can
also be constructed for the numerically obtained hoppings $b_n$ shown
in Fig.~\ref{fig_bs}.  The corresponding $|\phi|^2$ are plotted in the
lower panels.  Outside of free cases ($J_z = 0$), or integrable cases
($g = 0$), the generic behavior of the $b_n$, as shown in
Fig.~\ref{fig_bs}, will not host an exact edge state at the origin,
but rather only an approximate one.  This implies that $|\phi_l|^2$
will appear as a localized edge mode for small $l$, but for large $l$
will not be decaying sufficiently to yield a normalizable state.

The lower panels of Fig.~\ref{fig_bs} show that the decay of
$|\phi_l|^2$ is strongest at small $l$, corresponding to where the
staggering in the $b_n$ is the strongest.  Additionally, this initial
decay is more rapid for smaller $J_z$, which also coincides with $b_n$
that stagger with greater amplitude in comparison to the $b_n$ for
larger $J_z$. Moreover, the $b_n$ for smaller $J_z$ also stagger for a
longer range of $n$, i.e, the staggering continues further into the
bulk. As a result, for small $J_z$, non-trivial staggering of the
$b_n$ occurs beyond the end of the ramp, as opposed to that for larger
$J_z$. This is evident in the appearance of minima for $|\phi_l|^2$ at
small and intermediate $n$, for larger values of $J_z$, indicating
poor normalization of the mode.  In contrast, there is a lack of a
well defined minima for $|\phi_l|^2$ for smaller $J_z$, and thus a
behavior more akin to a SM.

We also note the differences between the $b_n$ obtained from the two
different numerical orthogonalization schemes in the top panels in
Fig.~\ref{fig_bs}. These differences do not affect the magnitude or
location of the minima of $|\phi_l|^2$ shown in the lower panels.

%%%%%%%%%%%%%%%%%%%%%%%%%%%%%%%%%%%%%%%%%%%%%%%%%%%%%%%%%%%%%%%%%%%%%%%%%%%%%%%%%%%%%%%%%%%%%
%%%%%%%%%%%%%%%%%%%%%%%%%%%%%%%%%%%%%%%%%%%%%%%%%%%%%%%%%%%%%%%%%%%%%%%%%%%%%%%%%%%%%%%%%%%%%
%%%%%%%%%%%%%%%%%%%%%%%%%%%%%%%%%%%%%%%%%%%%%%%%%%%%%%%%%%%%%%%%%%%%%%%%%%%%%%%%%%%%%%%%%%%%%
%%%%%%%%%%%%%%%%%%%%%%%%%%%%%%%%%%%%%%%
\subsection{Generalized SSH models}

The Heisenberg dynamics of $\sigma_1^x$ according to Eq.~\eqref{H}
generate $b_n$ that contain many details. However here we introduce a
series of toy models that address the main features of the $b_n$
addressed in Sec.~\ref{sec_3A}.

\begin{figure}
  \includegraphics[width = .49\textwidth]{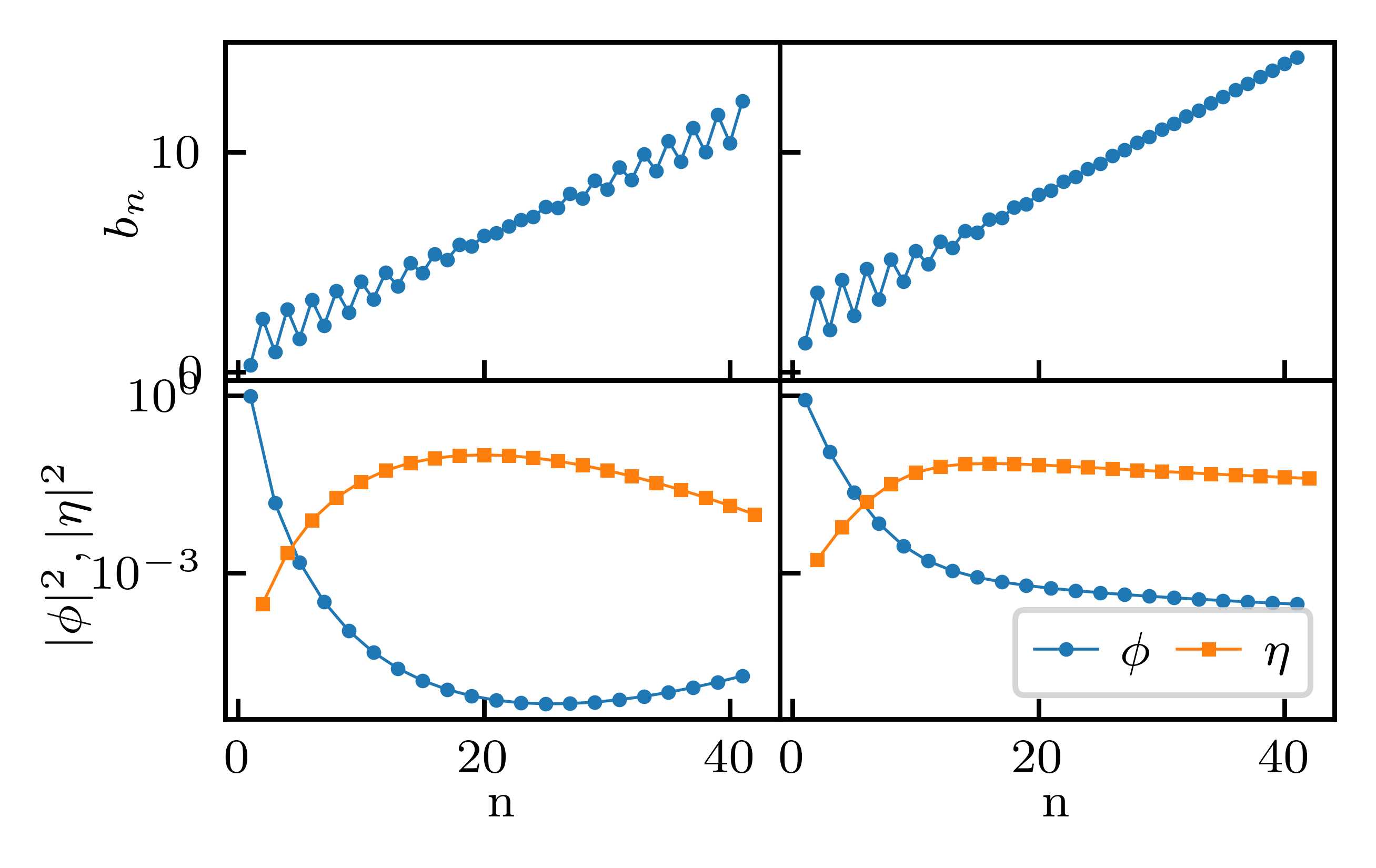}
  \caption{Demonstration of zero modes within the Krylov Hamiltonian.
    Top row: $b_n$ for the toy model Eq.~\eqref{toy1} on the left, and
    the toy model Eq.~\eqref{toy2} on the right.  Bottom row: the
    corresponding $|\phi|^2,|\eta|^2$ for the above $b_n$.  }
  \label{fig_toy1}
\end{figure}

The first model we consider is a nearest neighbor tight-binding model
with linear slopes for the hopping parameters, and with even and odd
site hopping parameters having different slopes,
\begin{equation}
b_n = \begin{cases} 
\alpha_1 n + \delta_1 & n \text{ odd}\\
\alpha_2 n + \delta_2 & n \text{ even}.
\end{cases}
\label{toy1}
\end{equation} 
An example of this model is plotted in the upper left panel of
Fig.~\ref{fig_toy1}.  We choose $\alpha_1 > \alpha_2$ and $\delta_1 <
\delta_2$ because this choice ensures that the even and odd hoppings
intersect for positive $n$. Moreover, in the $\alpha_{1,2}\ll 1$
limit, the region to the left of the intersection is topologically
non-trivial, with the intersection of the slopes equivalent to a
topological phase transition.  Consistent with this view, the edge
mode calculations for $\phi$ and $\eta$ respectively produce localized
states at the left boundary and at the topological transition. This is
shown in the lower left panel of Fig.~\ref{fig_toy1} where the
topological transition occurs around $n \sim 20$.  This figure also
shows that non-zero slopes $\alpha_{1,2}$ do not remove the edge-modes
of the SSH model entirely.

Now that we have shown that the zero mode considerations of the SSH
model are relevant, we move onto a second model, more aligned with the
ED data, in the thermodynamic limit.  In particular we consider
\(b_n\) that are such that the staggering falls off to zero in the
bulk, so that the bulk of the system can be thought of as a
metal. This model is,
\begin{subequations}\label{toy2}
\begin{align}
  b_n &= h_n + (-1)^n \tilde{h}_n, \label{eq_toy1}\\
  h_n &= \alpha n + \delta,\label{eq_toy2}\\
  \tilde{h}_n &= \frac{M_0}{2\left[\left( \frac{n}{n_0}\right)^\beta + 1\right]}.
  \label{eq_toy3}
\end{align}
\end{subequations}
Above, $M_0$ will have the interpretation of a mass, and we will show
this explicitly when we derive the continuum limit of the model.

An example of the above toy model is shown in the right column of
Fig.~\ref{fig_toy1}. The staggering of the $b_n$ starts off with a
maximum value of $M_0=2$ at the edge, and extends over a region of
width $n_0\sim 10$, before decaying to zero. The length scale over
which the staggering decays is $1/\beta = .25$. A linearly growing
hopping of rate \(\alpha=.3\) is superimposed on all the \(b_n\) and
we take $\delta=2$. We see in the bottom right panel of
Fig.~\ref{fig_toy1} that the $\phi$ zero mode remains (approximately)
localized, while the $\eta$ zero mode is no longer localized. This is
reminiscent of a topological phase transition where the edge mode
becomes delocalized when the phase is critical.  There is some decay
for both $\phi$ and $\eta$, well beyond $n_0$ which is slower than
$1/n$, and indicates that there are no longer any zero modes as the
wavefunctions are not normalizable.

The model in Eq.~\eqref{toy2}, after the mass has decayed, assumes
linearly growing $b_n$, i.e, a ramp that continues without
interruption. For a system of finite length however, this ramp will
eventually terminate into a plateau as seen in Fig.~\ref{fig_bs}.  In
order to compare with the numerical simulations of finite length
chains, we will consider $b_n$ which will follow Eq.~\eqref{toy2} up
to a certain distance, after which it will saturate and form a
"plateau" \cite{Barbon19}.  As we see saturation in the lifetime of
the $A_\infty$, we expect that as long as this plateau onset occurs
after the decay of the staggering, it should not affect the
lifetimes. This issue is also discussed in detail in Appendix
\ref{sec_lifetime_toy}.

Due to these considerations, we introduce another toy model, a
modification of Eq.~\eqref{toy2},
\begin{equation}
b_n = \begin{cases} h_n + (-1)^n\tilde{h}_n & n \le M\\
h_M & n>M
\end{cases},
\label{eq_toy4}
\end{equation}
where we take the point $M$ at which the ramp ends and the plateau
begins to be $M \gg n_0$. This toy model gives the same physics as the
model in Eq.~\eqref{toy2} at small $n$, but at $n = M$, the $b_n$
saturate in value, mimicking a metal.  This assumption of a perfect
flat metal for $h_{n>M}$ will serve as a good first order model.

%%%%%%%%%%%%%%%%%%%%%%%%%%%%%%%%%%%%%%%%%%%%%%%%%%%%%%%%%%%%%%%%%%%%%%%%%%%%%%%%%%%%%%%%%%%%%
%%%%%%%%%%%%%%%%%%%%%%%%%%%%%%%%%%%%%%%%%%%%%%%%%%%%%%%%%%%%%%%%%%%%%%%%%%%%%%%%%%%%%%%%%%%%%
%%%%%%%%%%%%%%%%%%%%%%%%%%%%%%%%%%%%%%%%%%%%%%%%%%%%%%%%%%%%%%%%%%%%%%%%%%%%%%%%%%%%%%%%%%%%%
%%%%%%%%%%%%%%%%%%%%%%%%%%%%%%%%%%%%%%%%%%%%%%%%%%%%%%%%%%%%%%%%%%%%%%%%%%%%%%%%%
\subsection{Comparison between the exact \(b_n\) and the toy model}

We now demonstrate that the above models will capture the essential
features of the exact $b_n$ obtained from the ED computation of the
Heisenberg dynamics of $\sigma_1^x$.

\begin{figure*}
  \includegraphics[width = .95\textwidth]{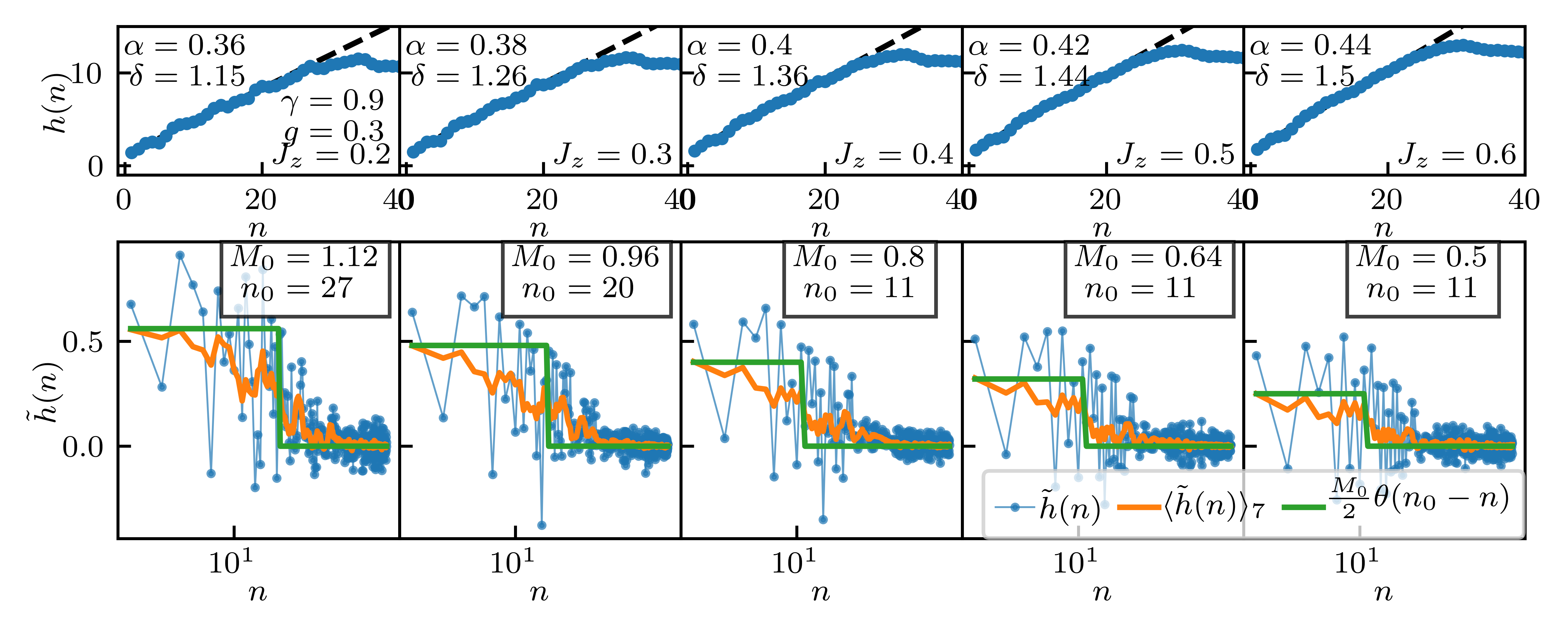}
  \caption{ $h_n,\tilde{h}_n$ plotted for $L = 12$ with $g = .3,\gamma
    = .9$ and several $J_z$, with numerical values determined for
    parameters $\alpha,\delta,M_0,n_0$. Top row: $h_n$ plotted along
    with a best fit of the linear ramp.  Bottom row: $\tilde{h}_n$
    plotted along with its moving 7-site average, $\langle
    \tilde{h}_n\rangle_7$, and with a square wave approximating the
    moving average $\theta(n_0 -n)M_0/2$.  $2 M_0$ is set as
    $\langle\tilde{h}_1\rangle_7$ and $n_0$ is determined starting at
    the far right $n$ and stepping backwards until $\langle
    \tilde{h}_n\rangle_7$ grows larger than $M_0$.}
  \label{fig_bn_fit}
\end{figure*}

With the aim of capturing the underlying trends proposed in
Eq.~\eqref{toy2} for small $n$, we consider neighboring averages and
differences as a simple numerical estimate of the parameters,
\begin{align}
h_n &\sim \frac{b_n + b_{n+1}}{2},\label{eq_num_h1}\\
\tilde{h}_n &\sim (-1)^n \biggl(\frac{b_n - b_{n+1}}{2}\biggr). \label{eq_num_h2}
\end{align}
The values for $h,\tilde{h}$ are plotted in Fig.~\ref{fig_bn_fit}.
Additionally, the best fit of the first twenty sites of $h_n$ is used
to extract the values of $\alpha,\delta$. While there is some small
variability between the different values of $J_z$, the best fit line
appears to accurately depict the $h_n$ values.

The $\tilde{h}_n$ plot is more complicated as it has a noise whose
amplitude is on the order of $M_0$.  A seven site moving average
allows for the trend of the mass to be better visualized. With the
averaging, the trend of $\tilde{h}_n$ does qualitatively agree with
the model of Eq.~\eqref{eq_toy3}. When $\beta \gg 1 $, the length
scale over which the mass drops to zero becomes smaller, and
Eq.~\eqref{eq_toy3} approaches a square wave function. A fit of the
$\tilde{h}_n$ to a square wave is also shown in the lower panel of
Fig.~\ref{fig_bn_fit}.

%%%%%%%%%%%%%%%%%%%%%%%%%%%%%%%%%%%%%%%%%%%%%%%%%%%%%%%%%%%%%%%%%%%%%%%%%%%%%%%%%%%%%%%%%%%%%
%%%%%%%%%%%%%%%%%%%%%%%%%%%%%%%%%%%%%%%%%%%%%%%%%%%%%%%%%%%%%%%%%%%%%%%%%%%%%%%%%%%%%%%%%%%%%
%%%%%%%%%%%%%%%%%%%%%%%%%%%%%%%%%%%%%%%%%%%%%%%%%%%%%%%%%%%%%%%%%%%%%%%%%%%%%%%%%%%%%%%%%%%%%
%%%%%%%%%%%%%%%%%%%%%%%%%%%%%
\section{Lifetime estimate from the continuum model}\label{Lanc}
%%%%%%%%%%%%%%%%%%%%%%%%%%%%%

We plan to derive an analytic estimate for the lifetime of the ASM of
the toy model in Eq.~\eqref{toy2}.  For this it is convenient to
derive a continuum version of this model. We outline the derivation
below with details relegated to Appendix \ref{app-cont}.

%%%%%%%%%%%%%%%%%%%%%%%%%%%%%
\subsection{Continuum model}

Let us write the toy model quite generally as the following
nearest-neighbor hopping Hamiltonian,
\begin{equation}
  H_K = \sum_n b_n ( c_n^\dagger c_{n+1} + c_{n+1}^\dagger c_n ).
\end{equation}
The corresponding Schrodinger equation takes the form,
\begin{equation}
  i \partial_t \Psi_n = b_n \Psi_{n+1} + b_{n-1} \Psi_{n-1}.\label{se1a}
\end{equation}
To transform to the continuum limit we assume that the hopping
parameters and the wavefunction can be written as
\begin{eqnarray}
  \Psi_n &=& i^n\left[\psi_n + (-1)^n \tilde{\psi}_n \right],
\label{psian} \\
  b_n &=& h_n + (-1)^n\tilde{h}_n,
 \label{bnan}
\end{eqnarray}
where $\psi_n,\tilde\psi_n,h_n,\tilde{h}_n$ are all assumed to be
smooth, slowly varying functions of $n$.

Measuring distance in lattice spacings $x=n$, introducing continuous
notations $\psi_n=\psi(x)$ etc., and in terms of the spinor,
\begin{equation}
  \tilde{\Psi} =
  \begin{pmatrix}
    \psi(x) \\ \tilde{\psi}(x)
  \end{pmatrix},
\end{equation}
the continuum
limit of Eq.~\eqref{se1a} becomes (see Appendix \ref{app-cont} for details),
\begin{equation}\label{cont10}
  i \partial_t \tilde{\Psi} =
  \left[ \sigma^y m(x) +
    \sigma^z\{i\partial_x,h(x)\} \right] \tilde{\Psi}.
\end{equation}

Above the mass is defined as,
\begin{equation}\label{m1def}
  m(x) = 2\tilde{h}(x) - \partial_x\tilde{h}(x).
\end{equation}

One can easily bring the Dirac equation \eqref{cont10} to the
conventional form using reparametrization of space coordinate. Namely,
changing variables $x\to X(x)$ so that
\begin{eqnarray}
    X &=& \int_0^{x} \frac{dx'}{2h(x')}\,, %{\color{blue}?? X = \delta \int_0^{x} \frac{dx'}{h(x')}}
 \label{Xtrans}\\
    \tilde{\Psi} &=& \frac{1}{\sqrt{h}}\chi\,,
 \label{psitrans}
\end{eqnarray}
we transform \eqref{cont10} to 
\begin{align}\label{cont20}
  i \partial_t \chi &= \left[\sigma^z i\partial_X +\sigma^y m(X)  \right] \chi\,,
\end{align}
where 
\begin{align}
    m(X) &= 2 \tilde{h}(X) -\frac{\partial_{X} \tilde{h}(X)}{2 h(X)} 
    \approx 2 \tilde{h}(X)\,.
 \label{htil}
\end{align}
In the last approximation we assumed that $\tilde{h}$ changes only
slowly in space which is consistent with the lattice toy model
\eqref{toy2}.

It is straightforward to construct a potential zero mode solution,
\begin{equation}
  {\chi}(X) = e^{\sigma^{x} \int_0^{X} m(Y)\, dY } {\chi}_0\,,
 \label{zm10}
\end{equation}
where ${\chi}_0$ is a constant spinor. The above is a time-independent
solution of the Schrodinger equation \eqref{cont10}. It corresponds to
a zero energy solution, and is a zero-mode or SM if normalizable.

We note the boundary condition of $\Psi(0) = 0$, translates into
$\sigma^x\chi(0)=-\chi(0)$ and ultimately into $\chi_0\sim (1,-1)^T$,
and so it is clear that the necessary condition for the normalization
of \eqref{zm10} is the condition $\int^X m(Y)\,dY\to +\infty$ as $X\to
+\infty$.  In particular the normalizability condition is satisfied
for $m(X)\to m>0$ and in this case Eq.~\eqref{zm10} is a true
zero-mode (or SM) solution of the Dirac equation \eqref{cont10} on a
half-line.

%%%%%%%%%%%%%%%%%%%%%%%%%%%%%%%%%%%%%%%%%%%%%%%%%%%%%%%%%%%%%%%%%%%%%%%%%%%%%%%%%%%%%%%%%%%%%
%%%%%%%%%%%%%%%%%%%%%%%%%%%%%%%%%%%%%%%%%%%%%%%%%%%%%%%%%%%%%%%%%%%%%%%%%%%%%%%%%%%%%%%%%%%%%
%%%%%%%%%%%%%%%%%%%%%%%%%%%%%%%%%%%%%%%%%%%%%%%%%%%%%%%%%%%%%%%%%%%%%%%%%%%%%%%%%%%%%%%%%%%%%
%%%%%%%%%%%%%%%%%%%%%%%%%%%%
\subsection{Connecting toy model parameters with the continuum model}\label{sec_conn}

Rewriting the toy model expression \eqref{eq_toy2} as $h(x)=\alpha
x+\delta$ and substituting it into \eqref{Xtrans} we obtain
\begin{equation}
    x=\frac{\delta}{\alpha}\left(e^{2\alpha X}-1\right)\,.
 \label{xX10}
\end{equation}
We can now proceed and express all parameters given by the toy model
\eqref{toy2} in continuum notations as functions of the transformed
coordinate $X$
\begin{align}
    h(X)  &= \delta e^{2\alpha X} \,,
 \\
    \tilde{h}(X) &=
    \frac{M_0}{2\left[ \left(\frac{e^{2\alpha X}-1}{e^{2\alpha X_0} -1} \right)^\beta+1\right]}\,,
 \\
    &\sim \frac{M_0}{2\left[e^{2\beta \alpha(X - X_0)} + 1\right]}\,,
 \\
    \Rightarrow m(X) &\sim \frac{M_0}{e^{2\beta \alpha(X - X_0)} + 1} \,.
 \label{toy-mass1}
\end{align}

We observe that in the toy model \eqref{toy2} the Dirac mass persists
for some range $x<x_0$ and then decays to zero. In the space
corresponding to the transformed coordinate $X$ the length of the
finite mass region is contracted exponentially. This makes sense as
the wavefunction/particle sees constantly increasing hopping strengths
as it progresses further into the bulk.

The equation \eqref{cont20} corresponds to a Dirac Hamiltonian with a
mass that is non-zero up to a certain distance \(\sim X_0\), beyond
which the Hamiltonian is gapless corresponding to that of a metallic
lead.

%%%%%%%%%%%%%%%%%%%%%%%%%%%%%%%%%%%%%%%%%%%%%%%%%%%%%%%%%%%%%%%%%%%%%%%%%%%%%%%%%%%%%%%%%%%%%
%%%%%%%%%%%%%%%%%%%%%%%%%%%%%%%%%%%%%%%%%%%%%%%%%%%%%%%%%%%%%%%%%%%%%%%%%%%%%%%%%%%%%%%%%%%%%
%%%%%%%%%%%%%%%%%%%%%%%%%%%%%%%%%%%%%%%%%%%%%%%%%%%%%%%%%%%%%%%%%%%%%%%%%%%%%%%%%%%%%%%%%%%%%
%%%%%%%%%%%%%%%%%%%%%%%%%%%%
\subsection{Decay rate in the continuum model}\label{sec_contdec}

Eq.~\eqref{toy-mass1} suggests that for large enough $\beta$, we may
replace the spatial dependence of the mass \eqref{toy-mass1} by the
step function
\begin{align}
    m(X) \approx M_0 \theta(X_0-X)\,.
 \label{toy-mass2}
\end{align}
The spectrum of the model \eqref{cont20} on a half-line $X\geq 0$ with
the mass \eqref{toy-mass2} is continuous. It has a quasi-stable mode
corresponding to an approximate zero mode \eqref{zm10}. The decay rate
of this mode can be found straightforwardly (see Appendix
\ref{sec_lifetime-cont} for details). It is given by
\begin{align}
    \Gamma_A \sim 4M_0 e^{-2M_0X_0}\,. 
 \label{gamma20}
\end{align}
The exponent can be clearly seen from the semi-classical expression
for tunneling amplitude
$\exp\left(i\int_0^{X_0}\sqrt{E^2-M_0^2}dX\Big|_{E=0}\right)=e^{-M_0
  X_0}$, corresponding to the tunneling probability
\eqref{gamma20}. Expressing \eqref{gamma20} in terms of the original
coordinate using \eqref{xX10} we obtain
\begin{align}
    \Gamma_A \sim 4M_0 e^{-\frac{M_0}{\alpha}
    \log\left(\frac{\alpha x_0}{\delta}\right)}\,. 
 \label{gamma100}
\end{align}

The decay rate \eqref{gamma100} is exponentially small in the size of
the staggering represented by $M_0$ in the continuum model. This
explains why the decay rate of the boundary spin is strongly
suppressed by the staggering in the SSH model in Krylov space. Both
the result \eqref{gamma100} and its derivation are very intuitive in
the continuum model (see Appendix \ref{sec_lifetime-cont}). However,
for the values of the parameters obtained from the spin chain (see
Fig.~\ref{fig_bn_fit}) the continuum limit of the Krylov Hamiltonian
is not fully justified. This is why in the next section we derive the
decay rate directly in the discrete model without appealing to
additional assumptions necessary for the continuum limit to hold.

%%%%%%%%%%%%%%%%%%%%%%%%%%%%%%%%%%%%%%%%%%%%%%%%%%%%%%%%%%%%%%%%%%%%%%%%%%%%%%%%%%%%%%%%%%%%%
%%%%%%%%%%%%%%%%%%%%%%%%%%%%%%%%%%%%%%%%%%%%%%%%%%%%%%%%%%%%%%%%%%%%%%%%%%%%%%%%%%%%%%%%%%%%%
%%%%%%%%%%%%%%%%%%%%%%%%%%%%%%%%%%%%%%%%%%%%%%%%%%%%%%%%%%%%%%%%%%%%%%%%%%%%%%%%%%%%%%%%%%%%%
\section{Lifetime estimate from the discrete toy model} \label{sec_green_setup}

We now calculate the lifetime of the ASM within the discrete setting
of the Krylov Hamiltonian. To this end, we consider a finite
Hamiltonian of length $N$, $H_{N}$, and connect it to a metallic
semi-infinite bulk.

In particular, the Hamiltonian \(H_N\) is a tight-binding model with
$N$ sites, and nearest neighbor hopping \(u_i\) and no onsite
potential.  We connect the right end of $H_N$ to the semi-infinite
metal, where the latter is modeled as a tight-binding model with
uniform hoppings. We are interested in the Green's function on the
first site, which we denote as the surface Green's function
$G_S^{(N)}(E)$.  Its explicit form is, (see Appendix \ref{disG}),
\begin{widetext}
\begin{equation}
  G_S^{(N)}(E) 
  = \left[(E - H_N -\hat{\Sigma})^{-1}\right]_{1,1} = \begin{pmatrix}
    E       & -u_1 &      &        &          &                \\
       -u_1 &E     & -u_2 &        &          &                \\
            &  -u_2&E     & \ddots &          &                \\
            &      &\ddots& \ddots &          &                \\
            &      &      &        &E         & -u_{N-1}       \\
            &      &      &        &-u_{N-1}  &E-\Sigma_{(N)}(E)
  \end{pmatrix}^{-1}_{1,1}.
  \label{eq_greens0}
\end{equation}
\end{widetext}
Above, the matrix $\hat{\Sigma}$ and its only non-zero element,
$\Sigma_{(N)}(E)$ in the lower diagonal, is the self-energy obtained
from integrating out the metallic lead.  In the rest of the paper we
always take $N$ to be even in order to avoid even-odd effects typical
of topological systems.  We also assume a smooth transition from $H_N$
to the metal by taking $u_N$ to be equal to the hopping amplitude of
the metal, i.e, $u_{N} = u_{N+1}$. This gives, (see Appendix
\ref{disG}),
\begin{align}
\Sigma_{(N)}(E) 
&= \frac{1}{2} 
\left(E - i \sqrt{4 u_{N}^2 - E^2} \right)\approx -i|u_N|,
\label{eq_sigma_1}
\end{align}
where in the last step we took the zero energy limit of the
self-energy. For this case the self-energy is purely imaginary
reflecting the fact that our system is open and the states decay into
the metallic bulk.

We are interested in solving for $G_S^{(N)}(E)$, when $H_N$ hosts an
approximate zero mode $\phi$ on the left end of the chain.  Whenever
the ASM $\phi$ is sufficiently strong and dominates the physics of the
first site, Eq.~\eqref{eq_greens0} can be solved in the small $E$
limit, yielding, (see Appendix \ref{disG} for details)
\begin{equation}\label{GSN1}
G_{(S)}^{(N)}
= \frac{(\mathcal{N}_\phi^{(N)} + R)^{-1}}{i\Gamma_A^{(N)}+E},
\end{equation}
where we define, 
\begin{subequations}\label{Ndef}
\begin{align}
 \mathcal{N}_\phi^{(N)} &= \sum_{l=1}^{N} |\phi_l|^2,\\
  \mathcal{N}_\eta^{(N)} &= \sum_{l=1}^{N} |\eta_l|^2,
\end{align} 
\end{subequations}
and,
\begin{equation}\label{Rdef}
R = \frac{|\phi_{N+1}|^4\mathcal{N}_\eta^{(N)} u_{N}^4}{\Sigma_{(N)}^2 u_1^2},
\end{equation}
with
\begin{equation}
    \Gamma_A^{(N)} = \frac{|\phi_{N+1}|^2 u_N^2}{ |\Sigma_{(N)}|\left( \mathcal{N}_\phi^{(N)} +R \right)}.
\label{eq_gammaA1}
\end{equation}
The forms of $|\phi_n|^2$, $|\eta_n|^2$ are derived from
Eq.~\eqref{eq_zero_modes}, and in particular we have, for $n = 2l+1$
\begin{equation}
    \phi_{2l+1}^2 = \prod_{k=1}^{l} \left[\frac{b_{2k-1}}{b_{2k}} \right]^2.
\label{eq_phi_sq}
\end{equation}
When $\phi$ is strongly localized, the quantity $R$ is vanishingly
small, $R\rightarrow 0$. For this case, performing the
Fourier-transform of Eq.~\eqref{GSN1}, we obtain,
\begin{align}
    A_\infty(t) &\approx \left( \mathcal{N}^{(N)}_\phi \right)^{-1} \exp\left[-\Gamma_A^{(N)} t\right]. 
\label{eq_ainf_approx1}
\end{align}

Note that, we can also numerically solve Eq.~\eqref{GSN1} on the real
axis and determine the decay rate from the half-width of the
Lorentzian of the imaginary part of the Green's function
$\Im\left[G(E\rightarrow 0)\right]$. This will yield the same decay
rate as in Eq.~\eqref{eq_gammaA1} when the edge mode is sufficiently
long lived.  Deviations from Eq.~\eqref{eq_gammaA1} will however
manifest when the lifetimes become shorter. This happens when $J_z$ is
larger, and since we are only interested in the limit of small $J_z$,
Eq.~\eqref{eq_ainf_approx1}, for our purposes is a good approximation
for the lifetime.

Below we discuss the lifetimes for two toy models. One corresponds to
an ideal SSH model coupled to leads. The other is the toy model of
Eq.~\eqref{eq_toy4} which is an SSH model with linearly growing
hopping amplitudes, and coupled to leads.

%%%%%%%%%%%%%%%%%%%%%%%%%%%%%%%%%%%%%%%%%%%%%%%%%%%%%%%%%%%%%%%%%%%%%%%%%%%%%%%%%%%%%%%%%%%%%
%%%%%%%%%%%%%%%%%%%%%%%%%%%%%%%%%%%%%%%%%%%%%%%%%%%%%%%%%%%%%%%%%%%%%%%%%%%%%%%%%%%%%%%%%%%%%
%%%%%%%%%%%%%%%%%%%%%%%%%%%%%%%%%%%%%%%%%%%%%%%%%%%%%%%%%%%%%%%%%%%%%%%%%%%%%%%%%%%%%%%%%%%%%
%%%%%%%%%%%%%%%%%%%%%%%%%
\subsection{ASM lifetime for ideal SSH model coupled to leads}

For the ideal SSH model, we take $H_N$ in Eq.~\eqref{eq_greens0} to be
one where $u_{2n-1} = u_{1}, u_{2n} = u_2$. Moreover, we take the
metal to have the hopping strength $u_0 = (u_1 + u_2)/2$.  The ratio
of the hoppings
\begin{equation}
r = \frac{u_1}{u_2},
\end{equation}
controls the topological nature of SSH model, where $r<1$ is
topological and $r>1$ is trivial.

In the $E \rightarrow 0$ limit, the surface Green's function can be
solved for exactly \cite{Zaimi19} (see Appendix \ref{disG} for
details), giving the decay rate,
\begin{equation}
%\Gamma_A \sim \frac{u_2^2 r^{N}(1 -r^2)}{|\Sigma(0)|}; \; |\Sigma(0)|=
\Gamma_A \sim \frac{u_2^2 r^{N}(1 -r^2)}{u_0} ;\;\;u_0=\frac{u_1+u_2}{2}.
\label{eq_lifetime}
\end{equation}

%%%%%%%%%%%%%%%%%%%%%%%%%%%%%%%%%%%%%%%%%%%%%%%%%%%%%%%%%%%%%%%%%%%%%%%%%%%%%%%%%%%%%%%%%%%%%
%%%%%%%%%%%%%%%%%%%%%%%%%%%%%%%%%%%%%%%%%%%%%%%%%%%%%%%%%%%%%%%%%%%%%%%%%%%%%%%%%%%%%%%%%%%%%
%%%%%%%%%%%%%%%%%%%%%%%%%%%%%%%%%%%%%%%%%%%%%%%%%%%%%%%%%%%%%%%%%%%%%%%%%%%%%%%%%%%%%%%%%%%%%
%%%%%%%%%%%%%%%%%%%%%%%%
\subsection{ASM lifetime for SSH model with slope and coupled to leads}

Let us estimate the decay rate \eqref{eq_gammaA1} for the toy model
Eq.~\eqref{eq_toy4}. Assuming the mass is a square wave of width $x_0$
and magnitude $M_0$ and using Eq.~\eqref{eq_phi_sq}, the decay rate is
found to be,
\begin{align}
  \Gamma_A&\propto|\phi_{x_0+1}|^2
  =\prod_{l=1}^{x_0/2}\left[\frac{\alpha(2l-1) + \delta - M_0/2}{\alpha(2 l) + \delta +M_0/2}\right]^2\,.
 \label{GammaAdiscr}
\end{align}
If $\alpha\ll M_0\ll \delta$ and $x_0$ is very large we evaluate
\eqref{GammaAdiscr} by replacing sums by integrals
and obtain with exponential accuracy
\begin{align}
    \ln \Gamma_A \sim -\frac{M_0}{\alpha}\ln\left(\frac{\alpha x_0}{\delta}\right)\,.
\end{align}
The above recreates the exponential dependence of the decay rate from
the continuum model, Eq.~\eqref{gamma100}.

%%%%%%%%%%%%%%%%%%%%%%%%%%%%%%%%%%%%%%%%%%%%%%%%%%%%%%%%%%%%%%%%%%%%%%%%%%%%%%%%%%%%%%%%%%%%%
%%%%%%%%%%%%%%%%%%%%%%%%%%%%%%%%%%%%%%%%%%%%%%%%%%%%%%%%%%%%%%%%%%%%%%%%%%%%%%%%%%%%%%%%%%%%%
%%%%%%%%%%%%%%%%%%%%%%%%%%%%%%%%%%%%%%%%%%%%%%%%%%%%%%%%%%%%%%%%%%%%%%%%%%%%%%%%%%%%%%%%%%%%%
\section{Comparison between ED and toy models} \label{comp}

\begin{figure}
\includegraphics[width = .48\textwidth]{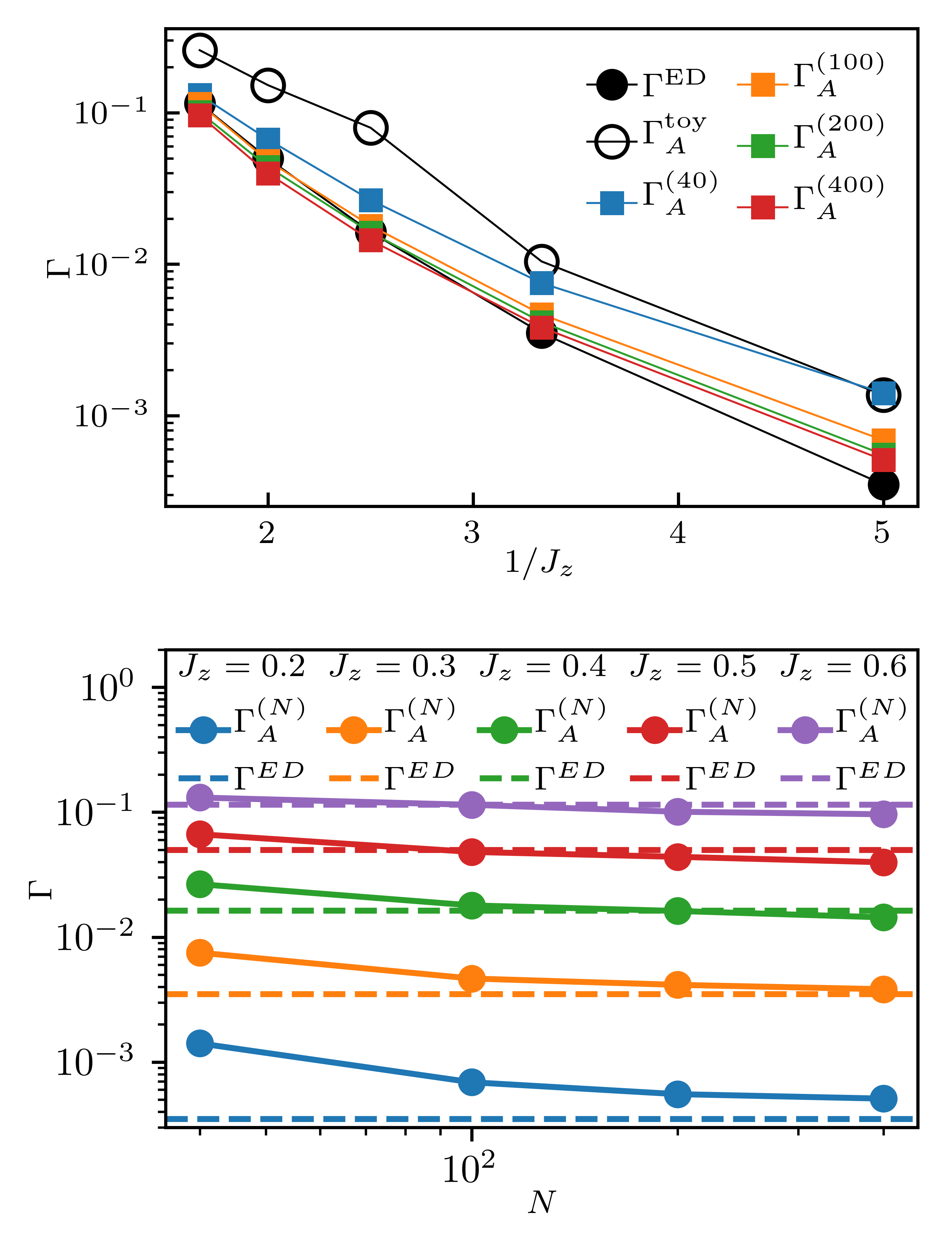}
\caption{ Upper panel: Comparison of the decay rates $\Gamma^{\rm ED}$
  obtained from ED to two approximate decay rates denoted by
  $\Gamma^{(N)}_A$ and $\Gamma^{\text{toy}}_A$.  The ED data set is
  obtained from the autocorrelation function already shown in
  Fig.~\ref{fig_ainfs}, with parameters $L = 12$, $g = .3,\gamma =
  .9$, and $J_z=.2,.3,.4,.5,.6$.  $\Gamma_A^{(N)}$ corresponds to
  Eq.~\eqref{eq_gammaA1} with $N=40,100,200,400$ and with the first
  $N$ Lanczos coefficients $b_n$ taken as the hopping parameters.
  $\Gamma^{\text{toy}}_A$ corresponds to Eq.~\eqref{eq_gammaA1}, but
  uses the toy model Eq.~\eqref{eq_toy4}, with the fitted values from
  Fig.~\ref{fig_bn_fit}, and taking $\beta = 100$, and $N = M = 50$.
  Lower panel: Convergence of $\Gamma_A^{(N)}$ to $\Gamma^{\rm ED}$,
  plotted against $N$ and for all $J_z$ values.  }
\label{fig_greens2}
\end{figure}

In this section we compare the lifetime determined from ED, to two
different estimates for the lifetime. One estimate is based on a
Krylov chain of length $N$, coupled to an ideal metal, where the
hoppings of the Krylov chain equal to the Lanczos coefficients in
Fig.~\ref{fig_bs}.  The second estimate is based on the toy model
Eq.~\eqref{eq_toy4}.

We now briefly return to Fig.~\ref{fig_ainfs}. This figure plots
$A_\infty$ from ED, and compares it to Eq.~\eqref{eq_ainf_approx1},
where the latter uses the $L=12$ Gram-Schmidt orthogonalized $b_n$,
and $N=400$. Eq.~\eqref{eq_ainf_approx1} accurately reconstructs the
prefactor of the decay, and is also a good approximation to the decay
constant. The latter is better visible in Fig.~\ref{fig_greens2}.  The
ED lifetime shown in Fig.~\ref{fig_greens2} is defined as $\Gamma^{\rm
  ED} = 1/t^*$, where $t^*$ is the time at which $A_\infty(t^*) =
(\mathcal{N}_\phi^{(N)} e)^{-1}$, where $\mathcal{N}_\phi^{(N)}$ is
given in Eq.~\eqref{Ndef}, with $N=400$.

The upper panel of Fig.~\ref{fig_greens2} compares the ED
lifetimes to the lifetimes $\Gamma_A^{(N)}$ determined from the
Lanczos coefficients and using the formula Eq.~\eqref{eq_gammaA1}.
The lifetimes $\Gamma_A^{\rm toy}$ obtained from the toy-model are
also shown, and these lifetimes are plotted against $1/J_z$.

In the computation of $\Gamma_A^{(N)}$, a few different $N$ have been
used. The dependence of $\Gamma_A^{(N)}$, as a function of $N$, shows
how the metallic approximation performs. Increasing $N$ brings
$\Gamma_A^{(N)}$ to within good agreement with $\Gamma^{\rm ED}$ for
intermediate values of $J_z$. The lower panel of
Fig.~\ref{fig_greens2} shows the convergence of $\Gamma_A^{(N)}$
towards $\Gamma^{\rm ED}$ as $N$ is increased.

There are two sources of error in Eq.~\eqref{eq_gammaA1} for
$\Gamma^{(N)}_A$. The first source of error is the negligence of
contributions that are higher order in $E$ (see Appendix
\ref{disG}). This error is manifested at larger $J_z$ values where, as
$N$ is increased, $\Gamma_A^{(N)}$ fails to saturate exactly on
$\Gamma^{\rm ED}$.  The second source of error is the failure to reach
values of $N$ that are large enough to account for the full decay of
the staggering in the $b_n$.  This source of error is visible for the
smaller $J_z$ data-sets.

In particular, the data set for smaller $J_z$ shows better agreement
between the lifetime obtained from ED and $\Gamma_A^{(N)}$ as $N$ is
increased, but some discrepancy remains after setting $N$ to the
largest value of $N=400$. In this case, staggering of the bulk $b_n$
extends far from the origin. This is also apparent in the behavior of
$|\phi_n|^2$ in Fig.~\ref{fig_bs}, where we see that $|\phi_n|^2$
fails to reach a minimum for the data set corresponding to smaller
$J_z$.

We now discuss the ability of the toy model Eq.~\eqref{eq_toy4} to
capture the essential physics.  In order to make the comparison, we
employ Fig.~\ref{fig_bn_fit} where the parameters of the toy model
Eq.~\eqref{eq_toy4} are extracted from the $b_n$ obtained from ED.
Moreover in the toy model we set $N=M=50\gg n_0$, and $\beta =100$ for
a sharp step.  Fig.~\ref{fig_greens2} shows the lifetime
$\Gamma_A^{\text{toy}}$ which is obtained from Eq.~\eqref{eq_gammaA1}
by using the parameters of the toy model.  For the data sets for
smaller values of $J_z$, the lifetime of the toy model agrees well
with $\Gamma_A^{(N)}$, for $N=40$.  For this case, both computations
for the decay rate make the same metallic approximation for the bulk
$b_n$, equivalently, both have no knowledge of any staggering present
in the bulk $b_n$. This agreement reflects that the step profile form
of the mass in the toy model is accurate for the longer lived ASM.

As previously mentioned, the deviation from the results from ED
indicates that the plateau of $b_n$ contains staggering that
contributes to the lifetime. This can be thought of as an additional
region of mass, which will lead to further decay of $|\phi_n|^2$ with
$n$. This additional decay, beyond the initial staggering, is the
strongest and most persistent for the smallest values of $J_z$, as
visible in Fig.~\ref{fig_bs}, where even the largest $L$, do not
appear to find a minima for the first $400$ $b_n$.  The remaining
$J_z$ data sets in Fig.~\ref{fig_bs}, do appear to see a bottoming out
for $|\phi|^2$, indicating that the first $400$ $b_n$ are sufficient
for accounting for any staggering present in the Lanczos coefficients
in the bulk.

%%%%%%%%%%%%%%%%%%%%%%%%%%%%%%%%%%%%%%%%%%%%%%%%%%%%%%%%%%%%%%%%%%%%%%%%%%%%%%%%%%%%%%%%%%%%%
%%%%%%%%%%%%%%%%%%%%%%%%%%%%%%%%%%%%%%%%%%%%%%%%%%%%%%%%%%%%%%%%%%%%%%%%%%%%%%%%%%%%%%%%%%%%%
%%%%%%%%%%%%%%%%%%%%%%%%%%%%%%%%%%%%%%%%%%%%%%%%%%%%%%%%%%%%%%%%%%%%%%%%%%%%%%%%%%%%%%%%%%%%%
%%%%%%%%%%%%%%%%%%%%%%%%%%%%%
\section{Conclusions} \label{conclu}

In this paper we have outlined a method to calculate the
non-perturbatively long lifetimes of edge modes that reside at the
boundary of non-integrable spin chains. Our approach is based on using
the recursion method to arrive at toy models, which can then be solved
analytically. We found good agreement with the lifetimes obtained from
ED and the analytic estimates from the toy models. We also have an
understanding for what causes the deviation between ED and toy
models. This arises primarily due to the ``ideal metal'' approximation
made for the bulk. In fact, the smaller is $J_z$, the more the bulk
deviates from this ideal metal limit, with staggering extending far
into the bulk.

Another possible source of discrepancy between ED and the toy models
is the presence of noise in the $b_n$s extracted from ED. As can be
readily seen from Fig.~\ref{fig_bn_fit}, the parameters $h_n$ and
$\tilde{h}_n$ are noisy both at small $n$ where staggering is present
and for large $n$ in the metallic bulk region.  The fluctuations of
$\tilde{h}_n$ at small $n$ correspond to mass fluctuations in the
continuous model. As the decay rate depends on mass exponentially, the
effects of this noise on the decay rate $\Gamma_A$ can be
essential. At large $n$ the noise in $h_n$ results in the suppression
of density of states near zero energy and might affect the tunneling
from the ASM into the metallic bulk as well. The study of these
effects is beyond the scope of this work.

While in this paper we presented results near the Ising limit ($\gamma
\sim 1$) of the non-integrable spin chain, our method is very
general. Future studies will explore the regime away from the Ising
limit, and will also apply it to Floquet ASMs\cite{Yates19}.  It is
also interesting to utilize this method to understand the stability of
edge modes in interacting topological insulators in higher spatial
dimensions.

%%%%%%%%%%%%%%%%%%%
{\sl Acknowledgements:} The authors thank Anatoly Dymarsky for helpful
discussions.  This work was supported by the US Department of Energy,
Office of Science, Basic Energy Sciences, under Award No.~DE-SC0010821
(DJY and AM) and by the US National Science Foundation Grant NSF
DMR-1606591 (AGA).

%%%%%%%%%%%%%%%%%%%%%%%%%%%%%
\appendix
%%%%%%%%%%%%%%%%%%%%%%%%%%%%%

%%%%%%%%%%%%%%%%%%%%%%%%%%%%%%%%%%%%%%%%%%%%%%%%%%%%%%%%%%%%%%%%%%%%%%%%%%%%%%%%%%%%%%%%%%%%%
%%%%%%%%%%%%%%%%%%%%%%%%%%%%%%%%%%%%%%%%%%%%%%%%%%%%%%%%%%%%%%%%%%%%%%%%%%%%%%%%%%%%%%%%%%%%%
%%%%%%%%%%%%%%%%%%%%%%%%%%%%%%%%%%%%%%%%%%%%%%%%%%%%%%%%%%%%%%%%%%%%%%%%%%%%%%%%%%%%%%%%%%%%%
\section{Numerical methods and orthogonalization errors in the $b_n$} \label{bnerror}

The numerical calculation of the $b_n$ can be efficiently performed by
representing the operators in their Pauli-string basis
\cite{Altman19}.  However, this choice of representation does not
avoid the usual costs associated with ED. In particular, for each
$|O_n)$ one needs to store the Pauli-string basis operators that
constitute $|O_n)$, and their non-zero coefficients. For
non-integrable systems, one expects the number of Pauli-string basis
elements in $|O_n)$ to increase exponentially with $n$. If one works
in the thermodynamic limit\cite{Altman19}, then this exponential wall
of complexity effectively caps the possible number of $b_n$ to roughly
$n \sim 40$.

While the thermodynamic limit is ideal for studying bulk properties
free of finite size effects, we choose to work with finite systems, of
the same size as the ED calculations, $L\le 14$. We are justified in
studying small systems because for the parameters chosen, $A_\infty$
saturates for the available system sizes, as seen in
Fig.~\ref{fig_ainfs}. By considering a finite system of size $L$, our
operators $|O_n)$ are bounded in length by $2^{2L}$, allowing us to
calculate more $b_n$.

For large $n$, the Lanczos algorithm is susceptible to errors in
producing orthogonal vectors.  Eq.~\eqref{eq_lanczos} requires only
three $|O_n)$ to be stored in memory at any given iteration step in
the algorithm. As numerical error accumulates, the newly generated
$|O_n)$ will inevitably have overlap with states calculated earlier,
and without storing those earlier vectors, there is no way to correct
for this. Typically this occurs between $n = 50$ to $n = 100$, which
is visible in Fig.~\ref{fig_bs}.

To correct for this orthogonalization error, one stores all $|O_n)$
calculated, and one performs Gram-Schmidt orthogonalization on the newly
generated states. Storing all calculated $|O_n)$, for large $n$, for
$L=14$, requires a large amount of memory, thus $L=12$ is the largest
system size for which we perform the extra Gram-Schmidt
orthogonalization step.

\begin{figure}
  \includegraphics[width = .5\textwidth]{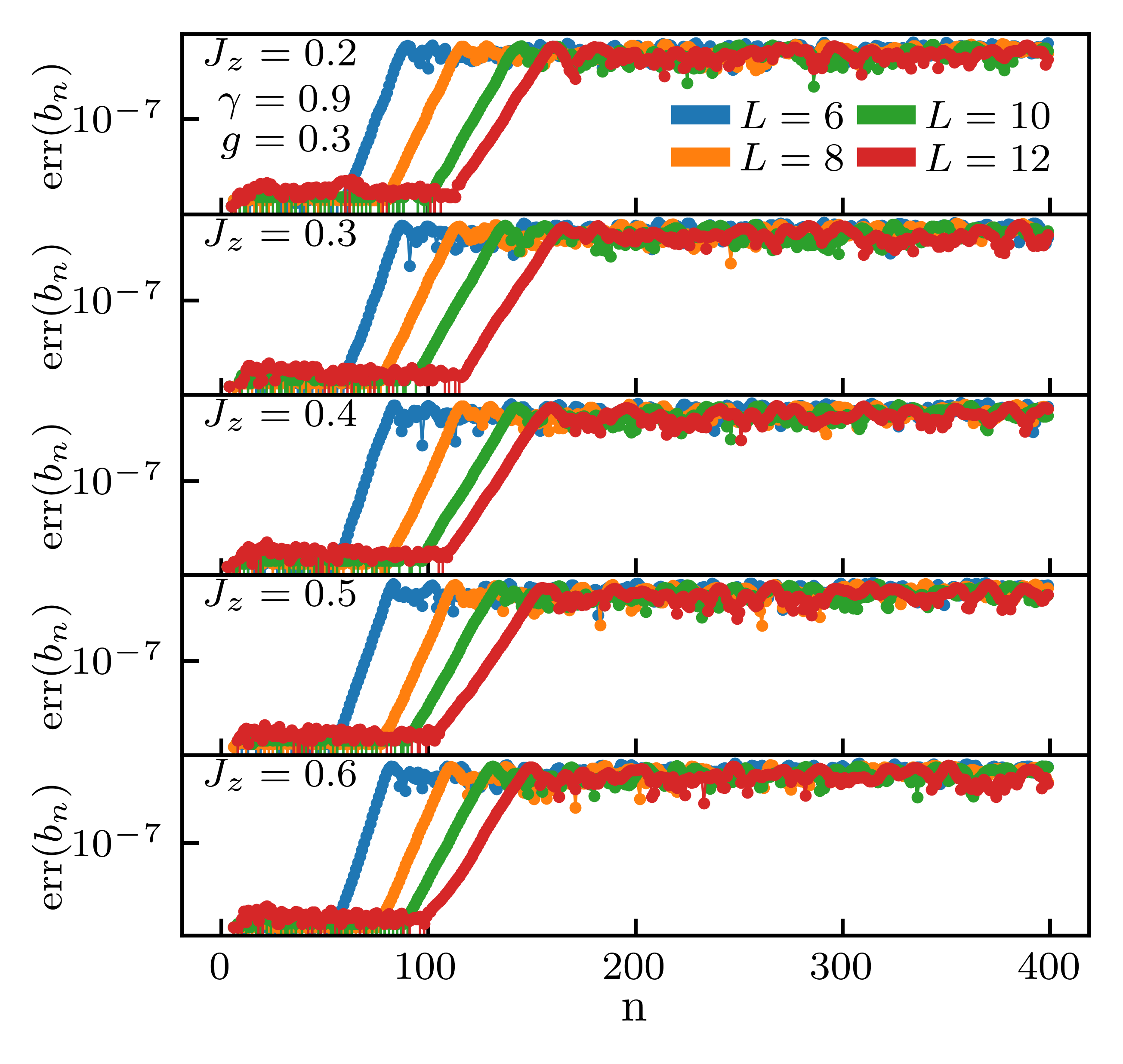}
  \caption{ Error in $b_n$ from performing the Lanczos algorithm
    without the additional Gram-Schmidt orthogonalization step, for $L
    = 6,8,10,12$ and $J_z = .2,.3,.4,.5,.6$. It is clear that the
    onset of the error increases in $n$ as one increases $L$, and that
    the rate of increase in error decreases as one increases $L$. The
    error also remains $\approx O(1)$ at large $n$.}
  \label{data_error}
\end{figure}

The differences between the $b_n$ obtained from the Lanczos algorithm
and the full Gram-Schmidt orthogonalization are shown in
Fig.~\ref{data_error}, for the available system sizes.  The error
saturates for large $n$ at a value of $\mathcal{O}(1)$.
Fig.~\ref{fig_bs} shows that the qualitative nature of the $b_n$ is
unchanged, with the most noticeable feature being the reduction of the
noise at large $n$ when the states are perfectly orthogonal.

%%%%%%%%%%%%%%%%%%%%%%%%%%%%%%%%%%%%%%%%%%%%%%%%%%%%%%%%%%%%%%%%%%%%%%%%%%%%%%%%%%%%%%%%%%%%%
%%%%%%%%%%%%%%%%%%%%%%%%%%%%%%%%%%%%%%%%%%%%%%%%%%%%%%%%%%%%%%%%%%%%%%%%%%%%%%%%%%%%%%%%%%%%%
%%%%%%%%%%%%%%%%%%%%%%%%%%%%%%%%%%%%%%%%%%%%%%%%%%%%%%%%%%%%%%%%%%%%%%%%%%%%%%%%%%%%%%%%%%%%%
\section{Derivation of the continuum Hamiltonian Eq.~\eqref{cont10}} \label{app-cont}

In Eq.~\eqref{se1a} we substitute the ansatz Eq.~\eqref{psian} for the
wavefunction, and the ansatz Eq.~\eqref{bnan} for the hopping
amplitudes to obtain,
\begin{widetext}
  \begin{align}
    &i \partial_t (\psi_n + (-1)^n \tilde{\psi}_n)
    = i\left[(h_n + (-1)^n\tilde{h}_n) (\psi_{n+1} - (-1)^{n} \tilde{\psi}_{n+1}) -
      (h_{n-1} - (-1)^n\tilde{h}_{n-1})(\psi_{n-1} - (-1)^n \tilde{\psi}_{n-1})\right]\nonumber\\
    &= i\left[
      h_n \psi_{n+1} - \tilde{h}_n \tilde{\psi}_{n+1}
      - h_{n-1} \psi_{n-1} - \tilde{h}_{n-1} \tilde{\psi}_{n-1}
      + (-1)^n(\tilde{h}_n \psi_{n+1} -h_n \tilde{\psi}_{n+1}
      + \tilde{h}_{n-1} \psi_{n-1} + h_{n-1} \tilde{\psi}_{n-1}
      )\right].
  \end{align}
  
  Since we are considering the wavefunctions and hopping parameters to
  be slow, while the $(-1)^n$ factor is rapid, we approximately solve
  the above by equating terms on the left and right to each other
  conditioned on the presence of the $(-1)^n$ term. This leads to the
  system of equations,
  \begin{align}
    i \partial_t \psi(n) &=
    i\left[ h_n \psi_{n+1} - \tilde{h}_n \tilde{\psi}_{n+1}
    - h_{n-1} \psi_{n-1} - \tilde{h}_{n-1} \tilde{\psi}_{n-1}\right],\\
    i \partial_t \tilde{\psi}(n) &=
    i \left[ \tilde{h}_n \psi_{n+1} -h_n \tilde{\psi}_{n+1}
    + \tilde{h}_{n-1} \psi_{n-1} + h_{n-1} \tilde{\psi}_{n-1}\right].
  \end{align}

  In order to take the continuum limit, we restore the lattice spacing,
  $n\rightarrow n a_0 = x$, and assume long wavelengths by dropping
  $\mathcal{O}(a_0^2)$ and higher order terms,
  \begin{align}
    i \partial_t \psi(x) &=
    i \left[ 2 a_0 h(x) \partial_x \psi(x) 
      -2 \tilde{h}(x) \tilde{\psi}(x)
      +a_0 \left[\partial_x h(x)\right] \psi(x)
      +a_0 \left[\partial_x \tilde{h}(x)\right] \tilde{\psi}(x) \right],\\
    i \partial_t \tilde{\psi}(x) &=
    i \left[ 2 \tilde{h}(x)\psi(x)
      - 2a_0 h(x)\partial_x\tilde{\psi}(x)
      -a_0 \left[\partial_x \tilde{h}(x)\right] \psi(x) 
      -a_0 \left[\partial_x h(x)\right] \tilde{\psi}(x)\right].
  \end{align}
\end{widetext}

By defining the spinor,
\begin{equation}
  \tilde{\Psi} =
  \begin{pmatrix}
    \psi(x) \\ \tilde{\psi}(x)
  \end{pmatrix},
\end{equation}
and the mass,
\begin{equation}
  m(x) = 2\tilde{h}(x) - a_0 \partial_x\tilde{h}(x),
\end{equation}
we have,
\begin{equation}
  i \partial_t \tilde{\Psi}(x) =
  \left[ \sigma^y m(x) + \sigma^z\{a_0 i\partial_x,h(x)\} \right] \tilde{\Psi}(x).
\end{equation}
We note that the boundary condition on $\tilde{\Psi}(x)$ is,
$\tilde\Psi(0) = 0 = \psi(0) + \tilde{\psi}(0)$, or,
\begin{equation}\label{bc}
  \sigma^x \tilde{\Psi}(0) = - \tilde{\Psi}(0).
\end{equation}

In what follows we find it convenient to set the lattice spacing
\(a_0=1\) to obtain Eq.~\eqref{cont10} in the main text,
\begin{equation}\label{cont2}
  i \partial_t \tilde{\Psi}(x) =
  \left[ \sigma^y m(x) +
    \sigma^z\{i\partial_x,h(x)\} \right] \tilde{\Psi}(x).
\end{equation}

%%%%%%%%%%%%%%%%%%%%%%%%%%%%%%%%%%%%%%%%%%%%%%%%%%%%%%%%%%%%%%%%%%%%%%%%%%%%%%%%%%%%%%%%%%%%%
%%%%%%%%%%%%%%%%%%%%%%%%%%%%%%%%%%%%%%%%%%%%%%%%%%%%%%%%%%%%%%%%%%%%%%%%%%%%%%%%%%%%%%%%%%%%%
%%%%%%%%%%%%%%%%%%%%%%%%%%%%%%%%%%%%%%%%%%%%%%%%%%%%%%%%%%%%%%%%%%%%%%%%%%%%%%%%%%%%%%%%%%%%%
\section{Lifetime in Dirac continuum model}
\label{sec_lifetime-cont}

Let us start with the Dirac equation on a half-line $x\geq 0$
\begin{align}
	i\partial_{t}\tilde{\Psi} &= H\tilde{\Psi}\,,
 \\
 	H &= \sigma^{z}i\partial_{x}+\sigma^{y} m(x)\,,
\end{align}
where the mass is space dependent. We assume that
$m(x)=m=\rm{const}>0$ for $0\leq x\leq x_{0}$ and $m(x)=0$ for $x\geq
x_{0}$. We also assume that the wavefunction is bounded at infinity,
and that at the origin $x=0$, the boundary condition is given in
Eq.~\eqref{bc}.

The wavefunction for $x<x_0$ is,
\begin{align}
\tilde{\Psi}(x<x_0) &= e^{-iEt}\left[ A_1 e^{-\kappa x}\left(\begin{array}{c} -i m\\i\kappa +E \end{array}\right)\right.\nonumber\\
&\left.	+ A_2 e^{\kappa x}\left(\begin{array}{c} -im \\-i\kappa + E\end{array}\right) \right]
\end{align}
while the wavefunction for $x\geq x_{0}$
\begin{align}
	\tilde{\Psi}(x\geq x_0,t) &= e^{-iEt}\left[ e^{-iE(x-x_{0})}\left(\begin{array}{c} 1\\0\end{array}\right)\right.\nonumber\\
&\left. 	+B e^{iE(x-x_{0})}\left(\begin{array}{c} 0\\1\end{array}\right) \right]\,.
\end{align}
Imposing boundary conditions at $x=x_{0}$ and $x=0$ we find the scattering amplitude $B$ to be
\begin{align}
	B &= - \frac{\kappa \cosh(\kappa x_{0})-m\sinh(\kappa x_{0})+i E\sinh(\kappa x_{0})}{\kappa \cosh(\kappa x_{0})-m\sinh(\kappa x_{0})-i E\sinh(\kappa x_{0})}\,,
\end{align}
where 
\begin{align}
	\kappa =\sqrt{m^{2}-E^{2}} \approx m-\frac{E^{2}}{2m}\,.
\end{align}
We used $|E|\ll m$ in the latter approximation. The scattering
amplitude $B$ has a pole in a complex plane of $E$ at (we assume
$mx_{0}\gg 1$):
\begin{align}
	E\approx -2im e^{-2mx_{0}} =-i\frac{\Gamma}{2},
\end{align}
giving the expression for the decay rate of the quasi-bound state:
\begin{align}
	\Gamma \approx 4m\, e^{-2mx_{0}}\,.
 \label{Gamma100}
\end{align}

Let us now note that the scattering amplitude $B$ is a pure phase $B=-e^{2i\delta}$ with 
\begin{align}
	\delta & =  \tan^{-1}\frac{E\sinh(\kappa x_{0})}{m e^{-\kappa x_{0}}-\frac{E^{2}}{2m}\cosh(\kappa x_{0})} \,.
\end{align}
We find that the derivative $d\delta/dE$ has a maximum at $E=0$:
\begin{align}
	\Delta t = 2\hbar \frac{d\delta}{dE}\Big|_{E=0} \approx \frac{\hbar}{m} e^{2mx_{0}}\,,
 \label{delaytime}
\end{align}
which should be interpreted as a time delay due to scattering. As
expected, the delay time \eqref{delaytime} is of the order as the
inverse decay rate \eqref{Gamma100}.

%%%%%%%%%%%%%%%%%%%%%%%%%%%%%%%%%%%%%%%%%%%%%%%%%%%%%%%%%%%%%%%%%%%%%%%%%%%%%%%%%%%%%%%%%%%%%
%%%%%%%%%%%%%%%%%%%%%%%%%%%%%%%%%%%%%%%%%%%%%%%%%%%%%%%%%%%%%%%%%%%%%%%%%%%%%%%%%%%%%%%%%%%%%
%%%%%%%%%%%%%%%%%%%%%%%%%%%%%%%%%%%%%%%%%%%%%%%%%%%%%%%%%%%%%%%%%%%%%%%%%%%%%%%%%%%%%%%%%%%%%
\section{Discrete Green's functions} \label{disG}
In this section we outline the steps needed to derive the results in
Section \ref{sec_green_setup}.

We first explain how the metallic bulk can be integrated out.  The
arguments here follow Ref.~\onlinecite{Zaimi19}, but we include the
steps for the convenience of the reader.  We define the metallic bulk
as a nearest-neighbor, tight-binding Hamiltonian $H_B$ with hopping
strength $u_0$ and no onsite potential.  The lattice Green's function
for the metallic bulk is defined as,
\begin{equation}
  G_{B} = (E - H_{B})^{-1}.
\end{equation}
We will only need the first component of this matrix, denoted as the
surface Green's function,
\begin{equation}
  G_{B}^S = (G_{B})_{1,1}.
\end{equation}
Using the identity,
\begin{equation}
  \left[\begin{pmatrix}
    A & B \\ C & D
  \end{pmatrix}^{-1}\right]_{1,1}\\
  = (A - BD^{-1}C)^{-1}\label{eq_mat_id},
\end{equation}
we can partition $H_B$ as follows,
\begin{equation}
  (E - H_B)^{-1} = 
  \begin{pmatrix}
    E & -u_0 \\ -u_0 & E - H_B
  \end{pmatrix}^{-1}.
  \label{eq_mat_2}
\end{equation}
Above the $(1,1)$ element is a scalar and the lower right element 
is semi-infinite. 
Solving for the $(1,1)$ element and using Eq.~\eqref{eq_mat_id}, we obtain,
\begin{equation}
G_{B}^S = (E - u_0^2 G_{B}^S)^{-1},
\end{equation}
whose solution yields,
\begin{equation}
G_{B}^S(E) = \frac{1}{2 u_0^2} \left(E \pm i \sqrt{4u_0^2 - E^2} \right).
\end{equation}
The sign above is chosen according to the initial conditions. For
example, a retarded Green's function will correspond to choosing the
lower sign.

If we introduce a qubit on the end of the metallic bulk
\cite{Zaimi19}, it will modify the Green's function as follows,
\begin{align}
  \left( E - H \right)^{-1}  &= 
  \begin{pmatrix}
    E - v_1 & -u_1 &  \\
    -u_1 & E-v_2 & -u_2  \\
    & -u_2 & E - H_B
  \end{pmatrix}^{-1},
\end{align}
where $H_B$ is semi-infinite. We can solve for the inverse of the top
left $2\times 2$ matrix using Eq.~\eqref{eq_mat_id}, giving the
following inverse Green's function for the qubit,
\begin{equation}
  \begin{pmatrix}
    E - v_1 & -u_1 \\ -u_1 & E - v_2 - u_2^2 G_B^S(E)
  \end{pmatrix}^{-1}.
\end{equation}
Due to the nearest-neighbor hopping, the effects of the bulk are
contained in an energy-dependent self-energy term on the last site,
\begin{equation}
\Sigma(E) = u_2^2 G_B^S(E).
\label{eq_self_eng_def}
\end{equation}
This motivates Eq.~\eqref{eq_greens0} in the main text. It also
explains the expression for the self-energy in Eq.~\eqref{eq_sigma_1}
where the hopping $u_2$ between the system and the metal has been
taken to be equal to the hopping in the metal $u_0$. Moreover the
latter is denoted by $u_N$ in the main text.

We now plan to solve Eq.~\eqref{eq_greens0} in the limit of $E
\rightarrow 0$.  We consider $H_{N}$ to have only nearest neighbor
hopping, with the hopping on site $n$ being $u_n$, and 
we exclude any onsite potential. 
$N$ will be taken to be even, thus $u_N$ will act
as the coupling between $H_{N}$ and the infinite bulk. Moreover, we
will treat the bulk as an ideal metal with uniform hoppings $u_{n \ge  N} = u_N$.

We denote $F_N$ as the lower right most matrix element of
Eq.~\eqref{eq_greens0}, inside the inversion. 
By utilizing
Eq.~\eqref{eq_mat_id}, 
we can solve
Eq.~\eqref{eq_greens0}, by iterating backwards from $F_N$ all the way
to $F_1$.
Let us denote an intermediate lower right matrix element as 
$F_n$, we then have, from Eq.~\eqref{eq_mat_id},
\begin{align}
F_{n-1} &= E - u_{n-1}^2 F_{n}^{-1}\nonumber\\
&= E - u_{n-1}^2 \frac{1}{E - u_n^2 F_{n+1}^{-1}}.
\end{align}
Above, in the last line we have used the fact that the same relation
holds between $F_n,F_{n-1}$ as between $F_{n+1},F_n$.   
We are eventually interested in the surface Green's function
$G_{(1,1)} = F_1^{-1}$. Below we outline the steps in the iteration.

Working to first order in $E \rightarrow 0$, and denoting $\Sigma =
\Sigma(E =0)$, we use that $F_N=E-\Sigma$. Then, iterating backwards
we obtain,
\begin{widetext}

\begin{equation}
\begin{split}
F_{N-(2l-1)} &\approx 
E + \frac{u_{N-(2l-1)}^2  \dots u_{N-3}^2}
{u_{N-(2l-2)}^2 \dots u_{N-2}^2}
\frac{u_{N-1}^2}{\Sigma}
\left[1 + E \left(
\frac{1}{\Sigma}
\left\{
\frac{u_{N-(2l-3)}^2 \dots u_{N-1}^2}{u_{N-(2l-2)}^2 \dots u_{N-2}^2} +
\frac{u_{N-(2l-5)}^2 \dots u_{N-1}^2}{u_{N-(2l-4)}^2 \dots u_{N-2}^2} +
\dots + 1 \right\} \right. \right. \\
& \left. \left. +
\frac{\Sigma}{u_{N-1}^2}\left\{
\frac{u_{N-(2l-4)}^2 \dots u_{N-2}^2}{u_{N-(2l-3)}^2 \dots u_{N-3}^2} + 
\frac{u_{N-(2l-6)}^2 \dots u_{N-2}^2}{u_{N-(2l-5)}^2 \dots u_{N-3}^2} + 
\dots + 1 \right\}
\right) \right].
\end{split}
\end{equation}

We now consider an $N = 2L$ subsystem $H_{N}$ and we
set $l = L$ in the above obtaining,
\begin{align}
G_{(1,1)} &\approx\left[
\left(\frac{u_1 u_3 \dots u_{N-1}}{u_2 u_4 \dots u_N} \right)^2 \frac{u_N^2}{\Sigma}+
E \biggl\{
\left( \frac{u_1^2\dots u_{N-1}^2}{u_2^2 \dots u_{N}^2} \right)^2
\frac{u_N^4}{\Sigma^2 u_1^2}
\left(1 + 
\frac{u_2^2}{u_3^2} + 
\dots +
\frac{u_2^2 \dots u_{N-2}^2}{u_3^2 \dots u_{N-1}^2}\right)
\right.  \nonumber \\ 
& \qquad  \left. +
\left( 1 + \frac{u_1^2}{u_2^2} + \dots + 
\frac{u_1^2 \dots u_{N-3}^2}{u_2 \dots u_{N-2}^2} \right)
\biggr\}\right]^{-1}\\
&= \left[
|\phi_{N+1}|^2 \frac{u_{N}^2}{\Sigma}+
E \left(|\phi_{N+1}|^4 \frac{u_{N}^4}{\Sigma^2 u_1^2}
\mathcal{N}_\eta^{(N)}+ \mathcal{N}_\phi^{(N)}\right)\right]^{-1}\\
&= \left[
|\phi_{N+1}|^2 \frac{u_{N}^2}{\Sigma}+
E \left(R+ \mathcal{N}_\phi^{(N)}\right)\right]^{-1},
\end{align}
where we have defined the wavefunction norms, 
$\mathcal{N}_\phi^{(N)} = \sum_{l = 1}^{N} |\phi_l|^2,\ 
\mathcal{N}_\eta^{(N)} = \sum_{l = 1}^{N} |\eta_l|^2$. 
We have also defined $R = |\phi_{N+1}|^4\mathcal{N}_\eta^{(N)} u_{N}^4/(\Sigma^2 u_1^2)$.

In its final form, the surface Green's function is, 
\begin{equation}
G_{(1,1)}
\approx \frac{(\mathcal{N}_\phi^{(N)}+R)^{-1}}
{|\phi_{N+1}|^2u_{N}^2 \Sigma^{-1}\left(\mathcal{N}_\phi^{(N)} + R\right)^{-1}+E}
\approx \frac{\left( \mathcal{N}_\phi^{(N)} \right)^{-1}}
{|\phi_{N+1}|^2 u_{N}^2 \left(\Sigma \mathcal{N}_\phi^{(N)} \right)^{-1}+E},
\end{equation}
where we have set $R=0$ in the second expression, a valid
approximation when $\phi$ is strongly localized.  From the last
expression above we can read off the lifetime as,
\begin{equation}
\Gamma_A^{(N)} \approx \frac{|\phi_{N+1}|^2 u_N^2}{|\Sigma| \mathcal{N}_\phi^{(N)}}.
\label{eq_lifetime_app}
\end{equation}

For a SSH model with homogeneous couplings coupled to a metal, we have
$u_{2l-1} = u_1$, $u_{2l} = u_2$, $r = u_1/u_2$, and $|\phi_{N+1}| =
r^{N}$. The square of the norm for the case where the SSH part is
sufficiently long is $\mathcal{N}^{-1}_\phi \approx 1 - r^2$.  The
resulting lifetime is,
\begin{equation}
\Gamma_{A}^{(N)} \approx \frac{r^N (1-r^2) u_2^2}{|\Sigma|},
\end{equation}
and is also reported in the main text.
\end{widetext}

%%%%%%%%%%%%%%%%%%%%%%%%%%%%%%%%%%%%%%%%%%%%%%%%%%%%%%%%%%%%%%%%%%%%%%%%%%%%%%%%%%%%%%%%%%%%%
%%%%%%%%%%%%%%%%%%%%%%%%%%%%%%%%%%%%%%%%%%%%%%%%%%%%%%%%%%%%%%%%%%%%%%%%%%%%%%%%%%%%%%%%%%%%%
%%%%%%%%%%%%%%%%%%%%%%%%%%%%%%%%%%%%%%%%%%%%%%%%%%%%%%%%%%%%%%%%%%%%%%%%%%%%%%%%%%%%%%%%%%%%%
\section{Independence of the lifetime on the plateau height of toy model \eqref{eq_toy4}} \label{sec_lifetime_toy}

\begin{figure}
  \includegraphics[width = .49\textwidth]{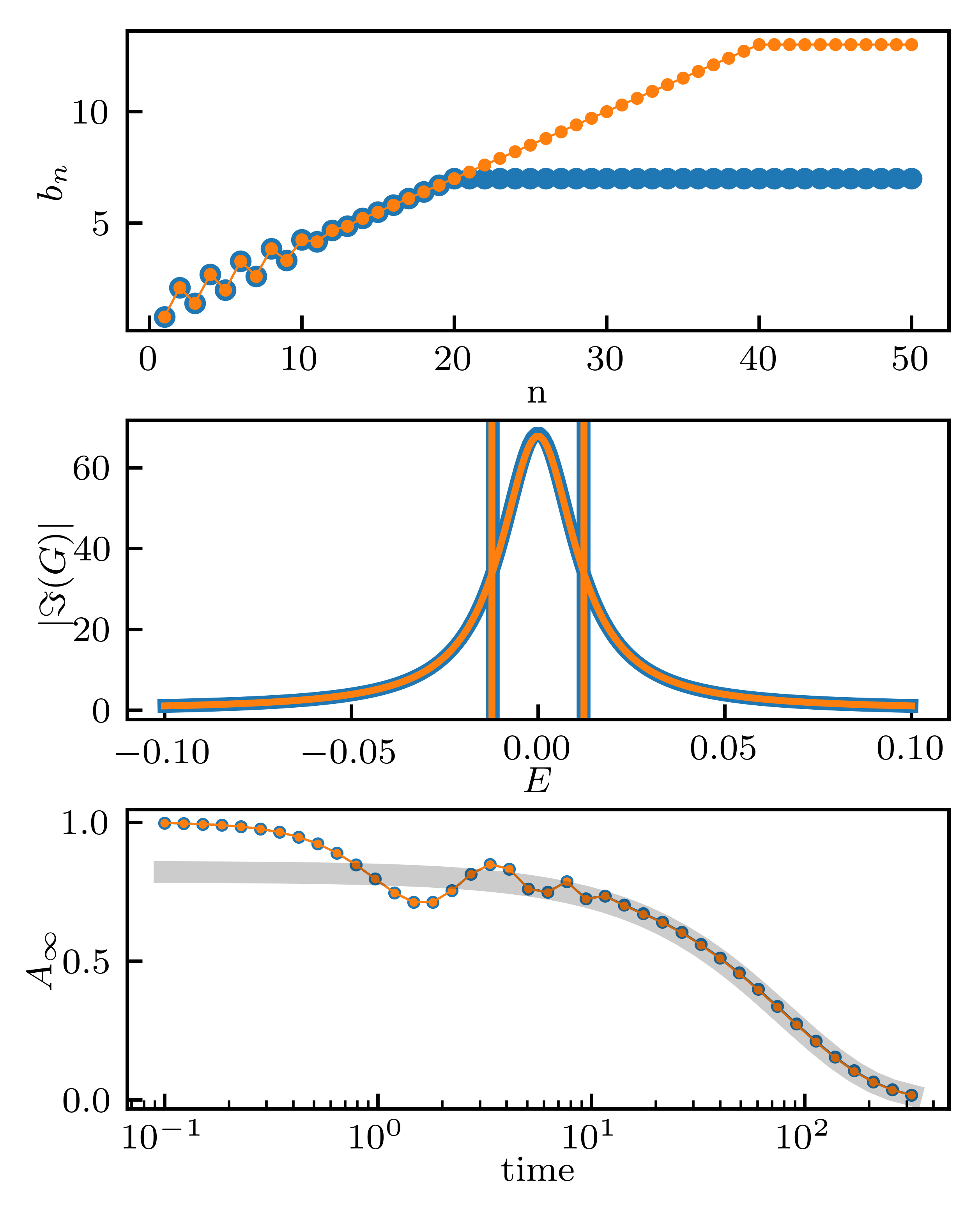}
  \caption{ Top panel shows two different configurations of the model
    in Eq.~\eqref{eq_toy4}, with the same parameters save for
    differing positions $M=20,40$ for the onset of the plateau.  The
    middle panel shows the imaginary part of the surface Green's
    function $|\Im G(E)|$ defined in Eq.~\eqref{eq_greens0}, with $N =
    50$ for both $M$.  The vertical lines in the middle panel mark the
    half-widths.  In the bottom panel, the colored dotted data sets
    show the time evolution on the toy model lattice for both models,
    where the plateau is made large, but finite, extending to 1e4
    sites.  The light grey thick line shows
    Eq.~\eqref{eq_ainf_approx1}, with $\Gamma^{(N)}_A$ corresponding
    to the half-widths shown in the middle panel.}
  \label{fig_toy_ramp_lengths}
\end{figure}

In this section we discuss the toy model of Eq.~\eqref{eq_toy4} with
the goal being to explain the dependence of the lifetime on the onset
of the plateau, whose position we denote by $M$.  We find that the
decay rate from Eq.~\eqref{eq_gammaA1}, when applied to the toy model
Eq.~\eqref{eq_toy4}, is insensitive to the the position $M$ at which
the plateau starts, so long as $M\gg n_0$, where $n_0$ is
approximately the position at which the staggering drops to zero.
This is also verified by numerically solving Eq.~\eqref{eq_greens0},
and demonstrated in Fig.~\ref{fig_toy_ramp_lengths}.

The top panel shows two different configurations of
Eq.~\eqref{eq_toy4} that correspond to two different values of $M$,
but share the following parameters $M_0=1,\beta=10, n_0=10,\alpha =
.3,\delta = 1$.  The resulting values of $|\Im G(E)|$, from
Eq.~\eqref{eq_greens0}, are plotted in the middle panel and are in
close agreement for the two different values of $M$.

The lower panel compares the time evolution on this lattice (colored
dotted lines) to that of Eq.~\eqref{eq_ainf_approx1} (grey thick
line). For the latter we use the decay-rate from the half-width of the
imaginary part of the Green's function, shown in the middle panel.
For the pre-factor of Eq.~\eqref{eq_ainf_approx1} we set $N =
50>M$. The lifetime is independent of $N$, as long as it is greater
than $M$.

In performing the numerical time-evolution (colored dotted lines in
lower panel), the system is initialized with the wavefunction
completely localized on the first site. The plateau is made finite,
but very large in order for the autocorrelation function of the first
site to be independent of the length of the plateau.  The calculation
is performed using iterative sparse matrix methods.  \cite{KrylovKit}

% \bibliography{zeromodes}


\end{document}